\documentclass[12pt,preprint]{aastex}%
\usepackage{epsf}
\usepackage{epstopdf}
\usepackage{graphics}

\newcommand{\NH}{\mbox{${\rm N}_{\rm H}$}} 
\newcommand{\AR}{\mbox{${\rm A}_{\rm R}$}} 

\newcommand{\xrtposra}{\mbox{RA(J2000)~=~21$^{\rm h}$22$^{\rm m}$9$\fs8$}}
\newcommand{\xrtposdec}{\mbox{ Dec(J2000)~=~$+77\degr$4$\arcmin$29\farcs0 }}
\newcommand{\xmmposra}{\mbox{RA(J2000)=21$^{\rm h}$22$^{\rm m}$9$\fs4$}}
\newcommand{\xmmposdec}{\mbox{Dec(J2000)=$+77\degr$4$\arcmin$28\farcs1 }}

\newcommand{\xmm}{{\it XMM-Newton }}
\newcommand{\swift}{{\it Swift }}

\newcommand{\degdot}{\mbox{$.\!^{\circ}$}}
\newcommand{\ls}
{\mathrel{\hbox{\rlap{\hbox{\lower4pt\hbox{$\sim$}}}\hbox{$<$}}}}
\newcommand{\gs}
{\mathrel{\hbox{\rlap{\hbox{\lower4pt\hbox{$\sim$}}}\hbox{$>$}}}}

\begin{document}
\title{GRB 050713A: High Energy Observations of the GRB Prompt and Afterglow Emission}
\author{D.C.~Morris\altaffilmark{1}, J.~Reeves\altaffilmark{2}, V.~Pal'shin\altaffilmark{3}, M.~Garczarczyk\altaffilmark{4},  A.D.~Falcone\altaffilmark{1}, D.N.~Burrows\altaffilmark{1}, H.~Krimm\altaffilmark{2}, N.~Galante\altaffilmark{5}, M.~Gaug\altaffilmark{6}, S.~Golenetskii\altaffilmark{3}, S.~Mizobuchi\altaffilmark{4}, C.~Pagani\altaffilmark{1}, A.~Stamerra\altaffilmark{5}, M.~Teshima\altaffilmark{4}, A.P.~Beardmore\altaffilmark{7}, O.~Godet\altaffilmark{7}, N.~Gehrels\altaffilmark{2}}

\altaffiltext{1}{Department of Astronomy \& Astrophysics, 525 Davey Lab., Pennsylvania State University, University Park, PA 16802, USA; {\it morris@astro.psu.edu}}
\altaffiltext{2}{NASA/Goddard Space Flight Center, Greenbelt, MD 20771, USA}
\altaffiltext{3}{Ioffe Physico-Technical Institute}
\altaffiltext{4} {Max-Planck-Institut f\"ur Physik, M\"unchen, Germany}
\altaffiltext{5} {Dipartimento di Fisica, Universit\`a  di Siena, and INFN Pisa, Italy}
\altaffiltext{6} {Institut de F\'\i sica d'Altes Energies, Barcelona, Spain}
\altaffiltext{7} {Department of Physics and Astronomy, University of Leicester, Leicester, LE1 7RH, UK}

\begin{abstract}
{\it Swift} discovered GRB~050713A and slewed promptly to begin observing with
its narrow field instruments 72.6 seconds after the burst
onset, while the prompt gamma-ray emission was still detectable in
the BAT. Simultaneous emission from two flares is detected in the
BAT and XRT. This burst marks just the second time that the BAT and
XRT have simultaneously detected emission from a burst and the
first time that both instruments have produced a well sampled,
simultaneous dataset covering multiple X-ray flares. 
The temporal rise and decay parameters of the flares are consistent
with the internal shock mechanism. In addition to the \swift coverage of
GRB~050713A, we report on the Konus-Wind (K-W) detection of the prompt
emission, an upper limiting
GeV measurement of the prompt emission made by the MAGIC
imaging atmospheric Cherenkov telescope and \xmm observations of the
afterglow. Simultaneous observations between \swift XRT and
\xmm produce consistent results, showing a break in the lightcurve
at T$_0$+$\sim$15~ks. Together, these four
observatories provide unusually broad spectral coverage
of the prompt emission and detailed X-ray follow-up of the afterglow
for two weeks after the burst trigger. Simultaneous spectral fits of K-W
with BAT and BAT with XRT data indicate that an absorbed broken powerlaw is often 
a better fit to GRB flares than a simple absorbed powerlaw. These spectral
results together with the rapid temporal rise and decay of the flares suggest
that flares are produced in internal shocks due to late time central engine activity.
\end{abstract}


\section{Introduction}

The {\it Swift} Gamma-Ray Burst Explorer \citep{Gehrels} has been
returning unprecedented data about gamma ray bursts (GRBs) for the
past 18 months. Of particular interest in the bursts followed by
\swift have been the early time lightcurves of the afterglows which have
shown much greater structure and different temporal decay properties
than expected, leading to much discussion in the literature
regarding the nature of the transition between the prompt emission, 
thought to be due to synchrotron radiation from internal collisions 
\citep{Gallant}, and the afterglow, also thought to be due to synchrotron 
radiation, though it remains somewhat unclear whether the emission 
arises in internal or external shocks. \swift observations have shown
that GRB lightcurves can be described by a canonical 3-segment shape. 
This shape consists of 1 - an early steep decay phase (F$_\nu \propto \nu^{-\beta} t^{-\alpha}$
where 3$\le$ $\alpha_1$ $\le$ 5; t $<$ 500~s) 2 - a very shallow decay phase
(0.5 $\le$ $\alpha_2$ $\le$ 1.0; 500~s $<$ t $<$ $10^4$~s) and 3 - a 'normal' decay
phase (1 $\le$ $\alpha_3$ $\le$ 1.5; $10^4$~s $<$ t) \citep{Nousek}.
Superimposed on this smooth
decay profile, \swift has shown that bright x-ray flares, 2 to 100
times as bright as the underlying afterglow, are common during the early (t $\le$ 10~ks) stages
of GRBs \citep{Burrows, Falcone, Piro, Romano}.

The observation of GRB050117 \citep{Hill} marked the first time that
\swift slewed to and settled on a GRB while the prompt gamma ray
emission was still in progress, arriving 192 seconds after the Burst
Alert Telescope (BAT) triggered on the burst which had a T$_{90}$ of
168 seconds. Due to irregularities in the
observing mode of the XRT and the proximity of the South Atlantic
Anomaly to \swift during the observation of GRB050117, however, only
very sparse data was collected by the XRT, totaling 11.4 seconds in 
the first orbit and 946 seconds overall.
This left large gaps in the lightcurve coverage and severely limited
the quality of the spectral analysis that could be performed.

We report here on the observation of GRB~050713A \citep{GCN3581}, a burst of T$_{90}$ = 70 
seconds to which \swift slewed and began collecting data
with the narrow field instruments (NFIs) in just 72.6 seconds, while
the prompt gamma ray emission was still detectable by the BAT. This
burst marks just the second time that the BAT and XRT have collected
simultaneous data on a burst and it marks the first time that both
instruments have produced a well sampled, simultaneous dataset
covering multiple flares in the prompt emission.

In addition to the {\it Swift} coverage of GRB~050713A, we report
also on prompt and followup observations carried out by Konus-Wind, 
MAGIC, \xmm and ground based optical observatories.
In section 2 we describe the observations and data analysis from all
instruments including ground follow-up. In section 3 we discuss the
implications of the observations in light of the new theoretical
understanding emerging from {\it Swift} observations of GRBs. In
section 4 we summarize and present our conclusions. Quoted uncertainties are
at the 90\% confidence level for one interesting parameter (i.e., $\Delta \chi^2$=2.71)
unless otherwise noted.

\section{Observations and Data Analysis}
Many different observatories and instruments have observed GRB~050713A.
We devote the following section to a description of the observations and 
analysis carried out by each instrument team. All spectral fits were performed using
XSPEC v11.3.

\subsection{\swift BAT}

The \swift BAT \citep{BAT} triggered on GRB~050713A at 04:29:02.39 UT, measuring a peak 1-second flux of $6.0 \pm 0.4$ \rm{photons cm$^{-2}$ s$^{-1}$}. T$_{90}$ measured in the 15--350~keV energy range is $70 \pm 10$~s \citep{Palmer}. The onset of the burst as defined by the BAT trigger is preceded by a weak, hard (photon index = 1.26) precursor at T$_0$--60~s. BAT data were processed using the BAT ground software build 11 and BAT Calibration Database files build 11.

At the time of the BAT trigger, the flux rose rapidly and remained elevated during a 12~s long, multipeaked burst (Fig 1). At T$_0$+12~s, the BAT flux rapidly decayed with a powerlaw decay rate of $\alpha\sim 8$ for 5 seconds before breaking to a more shallow decay of $\alpha\sim 2.5$ at T$_0$+17~s. This decay continued until T$_0$+40~s at which point the BAT flux had decayed to near background levels. At T$_0$+50~s, a flare is seen with peak flux $2 \times 10^{-8}$~ergs~cm$^{-2}$~s$^{-1}$, extrapolated into the XRT 0.2--10.0~keV bandpass, followed by a flare with peak flux $3.5 \times 10^{-8}$~ergs~cm$^{-2}$~s$^{-1}$ at T$_0$+65~s, another at T$_0$+105~s with peak flux $1 \times 10^{-8}$~ergs~cm$^{-2}$~s$^{-1}$ and some hint of further emission at the onset of a flare seen in the XRT at T$_0$+160~s. A weak but statistically significant precursor is seen at T$_0$--70~s to T$_0$--50~s followed by a period of no significant emission from T$_0$--50~s to the burst trigger.

The spectrum of the entire BAT dataset is well fit by a power-law spectrum with photon index =~$1.58 \pm 0.07$, though there is evidence for a slightly harder index of 1.45 during the plateau and a softening to $\Gamma=1.60$ during the rapid decay, and further softening to $\Gamma=2.0$ during the weak flares. Using the global fit of $\Gamma=1.58$, the fluence is $9.1 \pm 0.6 \times 10^{-6}$~ergs~cm$^{-2}$ in the 15--350~keV energy range.

\subsection{Konus-Wind}

GRB~050713A triggered Konus-Wind (K-W) \citep{Aptekar} at
T$_0$(K-W)= 04:29:01.745 UT. It was detected by the S2 detector,
which observes the north ecliptic hemisphere, with an incident angle of
18$\degdot$1. The K-W lightcurve in 3 bands is shown in Figure 2. The propagation delay from Wind to \swift is
1.387~s for this GRB. Correcting for this factor, one sees that
the K-W trigger time corresponds to T$_0$+0.742~s. Prior to 
T$_0$(K-W)-0.512~s data were collected by K-W in a survey 
mode with lower time resolution of 2.944~s and only 3 broad spectral
channels, 18--70~keV, 70--300~keV and 300--1160~keV.
From T$_0$(K-W) to T$_0$(K-W)+491.776~s,
64 spectra in 101 channels were accumulated on time
scales varying from 64 ms near the trigger to 8.19~s
by the time the signal became undetectable. The multichannel spectra cover the 18~keV--14~MeV energy range but
no statistically significant emission is seen above ~2~MeV. Data were processed
using standard Konus-Wind analysis tools.

Joint spectral analysis was carried out using the BAT data between 15 and 150~keV and the KONUS data from 20 to 2000~keV.
The spectra were fit by a power law model
with an exponential cut off:
dN/dE $\propto$ E$^{-\alpha}$ e$^{(-(2-\alpha)E/E_p)}$ where E$_p$
is the peak energy of the $\nu$ F$_\nu$ spectrum and $\alpha$ is the photon index.
The spectrum of the main pulse is well fit (Fig 3) with photon index $= 1.26 \pm 0.07$ and E$_p = 421_{-80}^{+119}$~keV ($\chi^2$=138/119 dof).
Joint fits between BAT and Konus
were also made for other time intervals, including one which shows the faint precursor detected by both
instruments at T$_0\sim$--60~s, and will be addressed in greater detail in
section 3.3.

The main pulse fluence in the 20~keV to 2~MeV range is
$8.08_{-1.77}^{+0.55} \times 10^{-6}$~erg~cm$^{-2}$.
The 256-ms peak flux measured from T$_0$+1.2~s in the 20~keV to 2~MeV band is
$1.34_{-0.45}^{+0.11} \times 10^{-5}$~erg~cm$^{-2}$~s$^{-1}$ and
the T$_{90}$ durations of the
burst in the G1, G2, G3 energy bands are $17 \pm 2$~s, $14 \pm 4$~s and $12 \pm 2$~s, respectively.

\subsection{\swift XRT}

The XRT \citep{XRT} performs an automated sequence of observations \citep{modes}
after \swift slews to a GRB detected by the BAT. When the spacecraft
first settles on the target, a short image (0.1~s followed by a
longer 2.5~s image if a position is not determined in 0.1~s) is taken
to determine an accurate position. Following the image, the XRT
switches into either Windowed Timing (WT) mode (a high timing
accuracy mode with 1 dimensional position information) if the source
count rate is above 2 counts s$^{-1}$, or Photon Counting (PC) mode (the more
traditional operating mode of X-ray CCDS in which full 2 dimensional
position information is retained but with only 2.5~s timing
resolution) if the count rate is below 2 counts s$^{-1}$.

XRT collected a 0.1~s Image Mode frame upon settling on GRB~050713A
73 seconds after the BAT trigger, which yields a count rate of 314
counts s$^{-1}$. Following the Image Mode frame, XRT cascaded down through
its automated mode sequence and collected its first WT frame 4.5
seconds later. At the onset of the WT data, the XRT count rate was
about 100 counts s$^{-1}$ and decaying as a powerlaw. This initial
powerlaw decay in the XRT WT data together with the Image Mode data
point measured at a flux level $\sim$3 times higher just 4.6~s earlier
clearly indicates that the XRT settled and began taking data during
the latter portion of the flare detected in the BAT at T$_0$+65 seconds
(see Fig 4).
XRT remained in WT mode throughout the entire first orbit of data
collection on GRB~050713A, also observing the flare detected by the
BAT at T$_0$+105 and a lower level flare not clearly detected by the BAT at
T$_0$+155~s.

Following a 65 minute period of occultation by the Earth, XRT began
observations again at T$_0$+4300~s, now observing in PC
mode since the countrate of the source had decayed below 2 counts s$^{-1}$.
A small flare at T$_0$+10~ks and the indication of another flare at T$_0$+45~ks are
seen in the late time XRT lightcurve data, superimposed on an otherwise steady powerlaw
decay. XRT observations continue to monitor the source until
T$_0$+1.8$ \times 10^{6}$~s, a total exposure time of 178~ks, at which time the source had decayed
below the XRT detection threshold.

XRT data are processed using the {\it xrtpipeline} ground software version 0.9.9, the redistribution
response matrices swxwt0to2\_20010101v007.rmf (WT data) and swxpc0to12\_20010101v007.rmf (PC data), and
ancillary response files generated with the {\it xrtpipeline} task {\it xrtmkarf}. 

\subsubsection{XRT GRB Position Analysis}

The X-ray afterglow position determined from ground processing of
the data is \linebreak \xrtposra \xrtposdec with
an uncertainty of 3.2 arcseconds. This is 10.5 arcseconds from the
reported BAT position, 0.5 arcseconds from the optical counterpart
reported by Malesani et al., \citep{Malesani} and 1.5 arcseconds from the initial XRT
position calculated onboard the satellite and automatically
distributed via the GCN network \citep{GCN3581}. An
X-ray image compiled from the first segment of XRT PC data is shown
as Figure 5 with the BAT, XRT and optical counterpart 
error circles displayed. A faint background source is detected 30 arcseconds due
south of the GRB afterglow at a constant flux level of $7 \pm 2 \times 10^{-4}$
counts s$^{-1}$. The contribution of this steady source has been removed from the
calculation of the afterglow lightcurve.

\subsubsection{XRT Temporal Analysis}

A timeline of the XRT (as well as other) observations of GRB~050713A is shown in Table 1.
The lightcurve will be broadly treated in two parts. The first part
is the initial orbit of data, during which the lightcurve is
characterized by bright flares which are simultaneously observed by
the BAT as well as the K-W instrument at higher energies. Due
to the extreme variability in this portion of the lightcurve, a
global decay index cannot be determined from the XRT data. The
second part is the remainder of the XRT data from the second orbit
onward, which is characterized primarily by a broken powerlaw decay,
though at least one small flare is seen
superimposed atop this global decay.

\subsubsubsection{First Orbit}

\swift finished slewing to GRB~050713A at T$_0$+73~s, during the flare
which began at T$_0$+65~s. The XRT short image frame is collected just 
after the peak of this flare, at a flux of $1.2 \times 10^{-8}$ergs~cm$^{-2}$~s$^{-1}$, 
and the first 20 frames of WT data record the decay of the flare. Fitting a
simple powerlaw to this decay from T$_0$+79~s to T$_0$+100~s setting T$_0$ to be the
BAT burst trigger time, we find a
powerlaw index of $5.6 \pm 1.8$ (1 $\sigma$). At T$_0$+105~s a new flare
begins, which rises with a powerlaw index of $23.3 \pm 4.5$ for
5-10~s, flattens at the peak of $\sim9 \times 10^{-9}$ ergs~cm$^{-2}$~s$^{-1}$ for 5-10~s, then decays
with a more shallow powerlaw index of $8.4 \pm 1.7$ for about 30~s.
At T$_0$+165~s a third flare is detected, which rises with a powerlaw
slope of $8.9 \pm 3.1$ for 5-10~s, flattens at the peak of $\sim1.5 \times 10^{-9}$ ergs~cm$^{-2}$~s$^{-1}$ 
for 5-10~s, then decays with a slope of $6.1 \pm 1.1$ for 70~s before the end of the observing window due to Earth
occultation.

\subsubsubsection{Second Orbit and Later}

The second orbit of data in the XRT is the only single orbit of data
in which the afterglow is characterized by a well sampled (greater
than 100 events total) lightcurve devoid of any obvious flaring activity.
During the 1600 seconds of data in this orbit, from T$_0$+4360~s to
T$_0$+5952~s, the lightcurve decays steadily as a powerlaw with decay
index of about 1.0. The third orbit of data is characterized by
another flare, beginning at T$_0$+10~ks, lasting throughout the entire
orbit (about 2~ks) and reaching a peak flux of $1 \times 10^{-11}$ ergs~cm$^{-2}$~s$^{-1}$. A powerlaw
fit to the rising portion of the flare yields a slope of $5.8 \pm 1.8$ 
while the decaying portion yields a slope of $11.0 \pm 2.5$.
This flare seems to be superposed atop the underlying afterglow decay of decay index $\alpha \sim 1$. 
Observations were interrupted after 150~s during the fourth orbit due
to the occurrence of GRB~050713B, and observations of GRB~050713A
remained suspended until T$_0$+40~ks. Some suggestion of another flare is
seen in the orbit of data beginning at T$_0$+45~ks, though the statistics
are poor. While afterglow data from the XRT alone do not clearly require 
a break in the afterglow powerlaw, \xmm data (see $\S$2.4) from T$_0$+21~ks to 
T$_0$+50~ks provide an accurate measure of the late-time decay slope ($\alpha=1.45$) which
cannot fit the XRT data from orbits 2 and 3 without a break in the powerlaw. The joint
XRT-\xmm lightcurve will be further discussed in $\S$2.4. 
Table 2 summarizes the flares and their temporal fits.

\subsubsection{XRT Spectral Analysis}

The XRT spectral analysis of this burst is somewhat complicated by
the high degree of flaring activity seen. 
In all cases, spectra are binned to a minimum of 20 counts per bin in order to 
use $\chi^2$ statistics. Fitting the entire first
orbit of data, the spectrum is well fit by a highly absorbed
powerlaw with photon index $=2.28 \pm 0.04$ and $\NH=4.8 \pm 0.2 \times 10^{21}$~cm$^{-2}$,
which is significantly
above the galactic column ($1.1 \times 10^{21}$~cm$^{-2}$) in the direction of
GRB~050713A. We are also able, due to the
large number of counts in each of the early flares in the dataset,
to fit a spectrum to both the rising and decaying portions of the
flares. In doing so we see the typical hard to soft evolution of the
flares \citep{Zhang}.

The second orbit of data shows a significantly different spectrum
from the first, with a harder spectrum with photon index $= 1.9 \pm 0.13$ 
and a lower value of $\NH=3.1 \pm 0.43 \times 10^{21}$, possibly indicating a period of
energy injection \citep{Nousek}. The third
orbit is well fit by a softer powerlaw similar to that which fit the
first orbit with photon index $=2.25 \pm 0.23$ and $\NH=4.1 \pm 0.7 \times 10^{21}$.

During the period of overlapping coverage between \swift and
\xmm, XRT has 3.5~ks of exposure time at a mean countrate of
0.04 counts s$^{-1}$ for a total of about 150 events during the
simultaneous observing period. Fitting a spectrum to this
overlapping coverage yields a photon index $=1.9 \pm 0.30$ and
$\NH=4.0 \pm 0.15 \times 10^{21}$. The corresponding mean unabsorbed
0.2--10.0~keV flux during the overlap period as measured by XRT is
$3.4 \pm 0.34 \times 10^{-12}$ ergs cm$^{-2}$~s$^{-1}$.

The data collected after the third orbit (i.e., after the temporal break in the lightcurve 
at T$_0$+$\sim20$~ks) are too sparse to justify fitting with
higher order models, but a simple absorbed powerlaw fit yields a spectrum of photon
index $= 2.8 \pm 0.6$ with $\NH=5.6 \pm 0.2 \times 10^{21}$. This is consistent with the x-ray 
photon index found in orbits 1 and 3 and is marginally softer than the photon index found during orbit 2
which, as noted above, suggests a period of energy injection.

\subsection{\xmm}

\xmm follow-up observations of GRB~050713A commenced T$_0$+23.6~ks
(for the EPIC-PN) and T$_0$+20.9~ks (for the two EPIC-MOS cameras). The \xmm\ 
data were processed with the \textsc{epproc} and \textsc{emproc}
pipeline scripts, using the \xmm\ SAS analysis package, version 6.5.
A bright rapidly decaying source is detected
near the aimpoint of all three EPIC detectors, localized at 
\xmmposra ~\xmmposdec. The net exposures after
screening and deadtime correction are 24.1~ks (PN) and 27.0~ks (MOS).
All three EPIC cameras (PN and 2 MOS) were used in Full Window Mode with
the medium filter in place.

Source spectra and lightcurves for all 3 EPIC cameras were
extracted from circular regions of 20 arcseconds radius centered on the
afterglow. Background data were taken from a 60 arcseconds
circle on the same chip as the afterglow, but free of any X-ray sources.
Fitting the afterglow
lightcurve with a simple power-law
decay results in a decay index of $\alpha=1.45 \pm 0.05$. 
Several flares are present in the background
lightcurve, so as a conservative check, we also excluded times where the
background rate is $>0.1$\,counts\,s$^{-1}$. The afterglow decay rate
is then $\alpha=1.39 \pm 0.09$, consistent with the above value. The decay rate
from the MOS lightcurve (for the two detectors combined) is also consistent
at $\alpha=1.35 \pm 0.06$. 

\subsubsection{\xmm Spectral Analysis}

Afterglow and background spectra were extracted with the same regions
used for the lightcurves, while ancillary and redistribution response files
were generated with the SAS tasks \textsc{arfgen} and \textsc{rmfgen}
respectively. As with XRT data, source spectra were binned to a minimum of 20 counts per bin
in order to use $\chi^{2}$ statistics.
The PN and MOS spectra were fitted jointly,
allowing only the cross normalization to vary between the detectors, which is
consistent within $<5\%$. The two MOS spectra and responses were
combined to maximize the signal to noise, after first checking that
they were consistent with each other. The average net source count rates
obtained over the whole observation are $0.58 \pm 0.01$\,counts\,s$^{-1}$
for the PN and $0.20 \pm 0.01$\,counts\,s$^{-1}$ per MOS module.

Allowing the absorption column to vary in the
spectral fit results in a formally acceptable fit
($\chi^{2}/{\rm dof}=515/496$). The $\NH$ obtained is
$3.1 \pm 0.1 \times 10^{21}$\,cm$^{-2}$, while the continuum photon index $=2.07 \pm 0.04$. 
The time-averaged, unabsorbed, 0.2--10.0~keV flux obtained for the
afterglow is $3.2 \times 10^{-12}$\,erg\,cm$^{-2}$\,s$^{-1}$. These values
are consistent with the \swift\ XRT measurement obtained at the time of the
\xmm\ observation.

The \xmm\ afterglow spectra were
also sliced into three segments of approximately 8~ks in length, in order
to search for any spectral evolution within the \xmm\ observation.
No change in the continuum parameters was found, all three spectral
segments being consistent with photon index $=2.1$ and
$\NH=3 \times 10^{21}$\,cm$^{-2}$. The spectrum obtained
from the PN detector and residuals to an absorbed power-law model
(with $\Gamma=2.08 \pm 0.02$ and $\NH=3.2 \times 10^{21}$\,cm$^{-2}$)
are shown in Figure 6. 

\subsubsection{Joint \xmm and \swift Modeling of the Late Time Afterglow}

The power-law decay index obtained from the \xmm\ observation
($\alpha=1.4$) appears
to be steeper than that obtained from the \swift\ XRT in orbit 2
($\alpha=1.0$). In order to compare
between the \xmm\ and \swift\ afterglow lightcurves, a combined lightcurve
from the \xmm\ and \swift\ observations was produced, scaling to the
absorbed continuum fluxes measured in the 0.5--10~keV band. The
joint \swift\ and \xmm\ lightcurve is shown in Figure 7, zoomed to better display 
the region at which the lightcurve break occurs.

A single power-law decay slope of $\alpha=1.20 \pm 0.02$
is an extremely poor fit to the lightcurve in this region, with a fit statistic of
$\chi^{2}/{\rm dof}=201.2/65$. Indeed the lightcurve from T$_0$+4~ks until
T$_0$+1000~ks can be better fitted with a broken power-law. 
There is a flat decay index of
$\alpha=1.02 \pm 0.07$ at early times and a steeper decay index of
$\alpha=1.45 \pm 0.06$ at later times, with the break in the decay occurring
at T$_0+25 \pm 3$~ks. The fit statistic is then
$\chi^{2}/{\rm dof}=90.2/59$. The remaining contribution
towards the $\chi^{2}$ originates from two small possible flares present near
T$_0+\sim10$~ks and T$_0+\sim45$~ks.

\subsection{MAGIC}

The MAGIC Telescope \citep{Mirzoyan} was able to observe part of the prompt emission phase
of GRB~050713A as a response to the alert provided by {\it Swift}.
The observation, at energies above $175\:\mathrm{GeV}$,
started at T$_0$+40~s, 20~s after reception of the alert.
It overlapped with the prompt emission phase measured by {\it Swift} and
K-W, and lasted for 37 min, until twilight.
The observation window covered by MAGIC did not, however, contain the
burst onset peak detected at keV-MeV energies, where the {\it Swift} and K-W
spectra were taken. The same region of the sky was observed 48 hours after
the burst onset, collecting an additional 49 minutes of data, which was used
to determine the background contamination.

\par

The MAGIC (\emph{Major Atmospheric Gamma Imaging Cherenkov\/}) Telescope
is currently the largest single-dish Imaging Air Cherenkov Telescope
(IACT) in operation, with the lowest energy threshold
($60\:\mathrm{GeV}$ at zenith, increasing with zenith angle).
In its fast slewing mode, the telescope can be repositioned
within $\sim$30~s.
In case of an alert by GCN, an automated procedure takes only a few seconds to
terminate any pending observation, validate the incoming signal and start slewing
toward the GRB position, as was the case for GRB~050713A.

\par

Using the standard analysis, no significant excess of $\gamma$-like air showers
from the position of GRB~050713A above $175\:\mathrm{GeV}$ was detected \citep{Albert}. 
This holds both for the prompt emission and during the subsequent observation
periods. Figure 8 shows the number of excess events during
the first 37 minutes after the burst, in intervals of 20~s. For comparison,
the number of expected background events in the signal region,
stable and compatible with statistical fluctuations, is shown.
Upper limits to the gamma-ray flux are given in Table 3.
This is the first observation of the GRB prompt emission phase performed
by an IACT.

\par

\subsection{Optical and Other Follow-up Observations}

Optical followup observations of GRB~050713A performed by the UVOT and by ground
based observatories are summarized in Table 4.

The earliest optical afterglow measurement comes from the RAPTOR-S 
robotic telescope at the Los Alamos National 
Laboratory in Los Alamos, New Mexico at R=18.4 $\pm$ 0.18 in a coadded series of 8x10 second images 
with a midpoint observation time of T$_0$+99.3~s \citep{Wren}. A nearly simultaneous measurement was made by the robotic
Liverpool Telescope in a coadded series of 3 $\sim$2 minute exposures in the r$'$ band with a midpoint of observation 
of T$_0$+3 minutes \citep{Malesani}.  Later detections below the Digitized Sky Survey limits were reported within the first 60 minutes after the
burst trigger in the R band by the Nordic Optical Telescope (T$_0$+47m), in the I band by the Galileo Italian National 
Telescope in the Canary Islands and in the infrared J,H, and K bands by the Astronomical Research Consortium 
Telescope at Apache Point Observatory (T$_0$+53m).  

Due to the bright (V=6.56) star HD204408 which is located just 68 arcseconds from the position of the burst, the UVOT
background level at the position of the afterglow is significantly higher than usual, resulting in abnormally 
poor sensitivity of the instrument in detecting the afterglow of GRB~050713A. Considering this high background, the 
non-detection of the afterglow by the UVOT is not surprising. 

All other reported optical observations of the afterglow position have yielded only upper limits. Most of the 
upper limits are near in time to the actual detections but at brighter magnitudes and thus do not produce 
strong constraints on the decay rate of the optical afterglow. The R-band measurement made at T$_0$+10.3 hours 
by the Lulin Telescope in Taiwan, however, is at a sufficiently late epoch to place a useful constraint on 
the optical decay rate. Fitting a simple powerlaw to the two well defined measurements at T$_0$+99.3~s and T$_0$+180~s 
and the upper limit at T$_0$+10.3 hours yields an upper limit on the power law decay slope of $\alpha \ge 0.5$, as is shown in Figure 4. 

A radio followup observation made with the VLA reports no detection at T$_0$+4.3 days.

\section{Discussion}

\subsection{Multispectral Lightcurve Overview}

The K-W light curve in the 18--1160~keV energy range is similar to the {\it Swift}-BAT
light curve (Fig 1).
The small precursor peak detected by BAT at T$_0$--70 to T$_0$--50~s
is seen by K-W at statistically significant levels in all three
broad, pre-trigger bands: G1 (18--70~keV), G2 (70--300), and G3 (300--1160~keV).
The other smaller peaks detected by the BAT after the burst trigger are not seen
at statistically significant levels in the K-W data,
despite the fact that the peaks at T$_0$+50~s and at T$_0$+65~s
are more intense in the BAT energy range than the precursor is.
The detection by K-W of the precursor but not the later flares is  
indicative of the harder spectral index seen in the precursor as 
compared to the later flares (see section 3.3 for discussion of
separate spectral fits to individual flares).

The XRT lightcurve with BAT data overplotted is shown in Figure 4. Both the X-ray and gamma-ray data in the first orbit are dominated by 
flaring activity, making it difficult to draw a conclusion regarding the underlying powerlaw decay index from this orbit alone.
The XRT data beginning at T$_0$+4~ks (orbit 2) and extending until T$_0$+40~ks show a significantly flatter powerlaw decay slope of $\alpha=-0.8$, implying
that a break in the powerlaw decay has occurred near the end of the first orbit of XRT coverage at T$_0$+$\sim$300~s and that a period of energy injection
occurs from T$_0$+$\sim$300~s to T$_0$+$\sim$15~ks. Another break in the lightcurve then occurs near T$_0$+25~ks to a steeper, 
``normal'', pre-jetbreak decay slope, as shown by the \xmm data ($\alpha\sim 1.4$).
Support for this notion of the presence of an energy injection phase may be drawn from the harder x-ray spectral slope of the second
orbit of XRT data (photon index = 1.9 $\pm$ 0.13) compared to the first orbit (photon index = 2.28 $\pm$ 0.04), the third orbit 
(photon index = 2.25 $\pm$ 0.23), and the later data (photon index = 2.8 $\pm$ 0.6) (Table 5). \xmm data 
coverage nicely fills much of the data gap in the XRT coverage between T$_0$+15~ks and T$_0$+40~ks and provides high signal to noise data in this 
regime, producing a confident determination of the lightcurve break. 

The global picture of the lightcurve of this burst is one in which the early data (prior to T$_0$+12~s) shows a bright plateau in the 15~keV 
to 1~MeV energy range, consisting of multiple overlapping peaks. At T$_0$+12~s the emission drops rapidly, consistent with a curvature radiation 
falloff \citep{Zhangb} until subsequent flaring activity begins to be seen in the 0.3--150~keV region with some indication of flux at higher energies 
from K-W. Due to the rapid rise and decay of 
the flares, internal shocks from continued central engine activity appears to be the most likely explanation for these flares \citep{Ioka}. 
The earliest ground based optical detections are reported at this time also, suggesting that the flares may also be optically 
bright. The lack of higher resolution timing information in the optical data, though, admits the possibility that the optical emission may be unassociated with the 
emission mechanism responsible for the x-ray flares. It is possible that the optical emission is due to synchrotron emission from the reverse shock (RS), 
though the much higher flux level of the x-ray flare peaks compared to the optical measurements suggests that the x-ray flares 
themselves are not due to inverse Compton scattering of the optical synchrotron emission of the RS \citep{Kobayashi, Gendre}. 

Following this 
prompt emission phase, an energy injection phase begins which dominates the lightcurve until at least T$_0$+16~ks. 
During the energy injection phase, continued activity of the central engine adds energy to the afterglow of the 
burst, either through additional ejection events or through the realization of energy contained in previously ejected outward moving relativistic 
shells which only collide at later times, producing late time internal shock emission which is then added to the overall decay \citep{Zhangb}.   
It may be expected, if the energy injection phase is due to continued central engine activity, that flaring behavior would continue to be
observed during this period and, indeed, some evidence for small scale flaring activity during both the second and third orbit of XRT data can be seen,
though at a much reduced significance in comparison to the flaring activity of the first orbit.
Near T$_0$+25~ks, the energy injection phase ends, giving way to a steeper decay slope similar to what is often seen in GRBs after the prompt emission 
phase and prior to the possible onset of a traditional jet-break \citep{Nousek}.

\subsection{Flares}

Many flares superimposed on top of the overall decay of GRB~050713A show the typical properties seen in other bursts: that $\delta$t/t $\sim$ 0.1 and that the 
peak flux level is negatively correlated with the time of the flare \citep{Falcone, Barthelmy}. These two properties of flares seen in
\swift GRB afterglows have been cited as evidence for flares being produced through accretion processes onto the central compact object 
\citep{Perna}, but we offer here that the constancy of the $\delta$t/t value of flares may partly be a by-product of the
overall decay of the afterglow in so much as the sensitivity of the XRT to flares is naturally degraded as the overall flux level 
of the afterglow decays, thus {\it requiring} flares at later times (and hence, lower flux levels) to be longer in duration for 
enough counts to be collected to produce a significant flare seen above the background. Such a case can be seen in comparing the 
early time flares in the first orbit of GRB~050713A to the flare seen in the the third orbit. During the first orbit, the underlying 
flux level beneath the flares is poorly determined, but can be assumed to be 10-100 counts s$^{-1}$. We are dominated in this portion of the
lightcurve by the Poissonian error in the flux, which in a 10 second integration will be 10-32 counts, or 3-10$\%$ . 
Thus, for a flare to appear at the 6 sigma level 
above the background during this portion of the lightcurve, at most a 60$\%$ increase in fluence above the normal powerlaw decay is needed, 
which can be acquired in a few seconds by the introduction of a flare with twice the flux of the underlying afterglow. 
During the third orbit, however, the underlying afterglow flux level
has dropped to $\sim$ 0.1 counts s$^{-1}$. During a 10 second integration at 0.1 counts s$^{-1}$ the Poisson error alone is 1 count, so for a flare to be 
detectable at 6 sigma above background at these count levels, the total fluence must be 6 counts, implying an increase in the rate from 0.1 counts s$^{-1}$ 
to 0.6 counts s$^{-1}$ during the 10~s interval, a 6 fold increase, which
has been seen only in the brighter flares. In order to be sensitive to the same 60$\%$ increase in flux level as during the
first orbit, the flare which occurs at a flux level of 0.1 counts s$^{-1}$ needs to have a Poission error which is 1/6 of the total counts in the
observation, i.e., 36 counts must be collected, which implies an exposure time of at least 180~s if produced by the introduction of a flare with twice the flux 
of the underlying afterglow. In other words, because the afterglow flux level decays as t$^{-\alpha}$, the exposure time needed to acquire the
same fluence level increases as t$^{\alpha}$. Thus, we see that in moving from the
first orbit at T$_0$+100~s to the third orbit at T$_0$+10000~s, assuming a typical underlying powerlaw decay of the afterglow of $\alpha \sim$ 1, 
we have greatly decreased the temporal resolution of XRT to detect flares (from a few seconds to a few hundred seconds). This is not to imply that there is not another more physical 
cause for the constancy of the $\delta$t/t ratio seen in flares, but rather to note that the typical GRB seen by the \swift XRT does not provide
sufficient flux at times typically greater than a few ks to detect the shorter timescale flares that are so often seen during the first orbit.

In GRB~050713A, a hint of emission above the afterglow powerlaw decay appears in the XRT data at T$_0$+45~ks, though the statistics are, predictably, poor. This time is overlapped by \xmm data, though, so we can look for evidence of a short flare in the \xmm data at this time. In Figure 9 we show the \xmm lightcurve, plotted linearly and zoomed near T$_0$+45~ks. Though a 1-2 sigma deviation above the background decay is seen at T$_0$+45~ks, the \xmm data appear consistent with a statistical fluctuation rather than a true flare similar to those seen earlier during the burst. 

The presence of multiple flares in GRB050713A argues against ``one-shot'' emission mechanisms such as synchrotron self-Compton emission in a reverse
shock or deceleration of the blastwave \citep{Piroa} and it argues in favor of a mechanism which can produce repeated flares, such as late time central
engine activity. While it may remain possible that one of the several flares in GRB050713A is due to the RS or the onset of the afterglow due to 
external shocks, the steep temporal decays of all the temporally fitted flares coupled with the photon indices of the flares (1.25 $\sim$ 2.5; Table 6) do not satisfy the closure relations of Sari et al., (1998), Chevalier \& Li (1999) and Sari et al., (1999) for propagation of the blast wave
into either a wind or constant density ISM. Together these points seem to argue in favor of an internal shock origin for the flares seen in this burst.

\subsection{Joint Spectral Fitting}

Due to the relatively narrow spectral response function of the BAT (15--150~keV for mask-tagged events) and the XRT (0.3--10~keV), 
a spectral fit to data from only one of the two high energy instruments on \swift is usually not able to discriminate between 
higher order spectral models. Analysts and authors are usually limited to choosing between a power-law or Band function. In
GRB~050713A we have a rare case of data coverage overlap between BAT and XRT (0.3--150~keV) and also between BAT and K-W (15~keV--14~MeV). Taking advantage of this where appropriate, considering the relative flux levels in the three instruments, we have jointly fitted spectral datasets between the two pairs of instruments. During the precursor and from T$_0$+0 to T$_0$+16.5~s, we perform joint fitting between BAT and K-W data. From T$_0$+16.5 to T$_0$+78~s we have only BAT data. From T$_0$+78 to T$_0$+116~s and during the onset of the flare at T$_0$+160 we perform joint fitting between XRT and BAT.
We have grouped the data into segments (as shown in Table 6) in order to temporally separate data which we expect may show significantly different spectral parameters. Segments 1-4 contain BAT and K-W data and are segmented to separate the precursor from the prompt emission and the prompt emission from the rapid decay phase. Segment 5 contains BAT data only and segments 6-10 contain XRT and BAT data. These are segmented to distinguish the 3 flares which have overlapping data and also to separate the rise of each flare from the decay of each flare. We attempt fits to each of these segments using 4 different spectral models: 1) an absorbed powerlaw 2) an absorbed cutoff or broken powerlaw (cutoff for data extending beyond 150~keV, broken otherwise) 3) an absorbed Band function and 4) an absorbed blackbody plus powerlaw.

\subsubsection{Segment 1: precursor (T$_0$--65 to T$_0$--55~s)}

The precursor is the most poorly sampled of all the regions. Despite the low number of counts in the region, a cutoff powerlaw is favored over a single powerlaw at $90\%$ confidence according to the F-test. Of all the segments fit, the precursor has the hardest photon index, regardless of the model which is used to perform the fits.

\subsubsection{Segment 2: prompt emission plateau (T$_0$+0 to T$_0$+8.5~s)}

The plateau of the prompt emission is best fit by an exponentially cutoff powerlaw model with photon index $=1.26$ and E$_{peak}=421$~keV. Next to the precursor, the prompt plateau has the second hardest photon index of all segments fit, regardless of the model used.

\subsubsection{Segment 3: rapid decay (T$_0$+8.5 to T$_0$+25~s)}

As with the other data segments which contain K-W data, the rapid decay segment is poorly fit by a simple powerlaw and is best fit by a cutoff powerlaw or Band function. The photon index of the cutoff powerlaw in segment 3 is quite similar to that in the prompt plateau, but the cutoff energy is somewhat lower (312~keV compared to 421~keV in the plateau), suggesting that the highest energy flux is ``shutting off'' during the rapid decay phase.

\subsubsection{Segment 4: plateau + early rapid decay (T$_0$+0 to T$_0$+16.5~s)}

This segment is an extension of the prompt segment to slightly later times, encompassing slightly more data. The cutoff powerlaw or Band function is the best fit, with photon indices similar to segment 2 and E$_{peak}$ between that in segments 2 and 3. 

\subsubsection{Segment 5: rise of T$_0$+60~s flare (T$_0$+59 to T$_0$+68~s)}

This segment contains only BAT data and is included for completeness, though the narrowness of the BAT spectral response limits the ability to distinguish between models. A simple powerlaw is a good fit with photon index of 1.83. \NH~is unconstrained. The powerlaw plus blackbody model produces a good fit to this segment but only with a very minimal blackbody component, effectively reproducing the fit of the simple absorbed powerlaw. Therefore we consider the powerlaw plus blackbody model inapplicable to this segment.

\subsubsection{Segment 6: decay of T$_0$+60~s flare (T$_0$+68 to T$_0$+95~s)}

Only in this segment, the data time ranges are mismatched between XRT and BAT (due to XRT observations beginning towards the end of the flare decay). Rather than ignore this flare or consider only the later part of the flare decay where XRT and BAT data coverage overlap, we have chosen to fit the entire BAT time range from T$_0$+68 to T$_0$+95~s together with the T$_0$+79 to T$_0$+95~s XRT data (note that the Image mode data taken at T$_0$+73~s are highly piled up and cannot be used spectrally) for consistency with our treatment of the other flares. A simple powerlaw is a good fit to this segment, yielding $\NH=5.2 \times 10^{21}$cm$^{-2}$ and a photon index of 2.47, significantly softer than the rise of the flare, as expected.

\subsubsection{Segment 7: rise of T$_0$+100~s flare (T$_0$+100 to T$_0$+113~s)}

In the rise of the brightest flare seen in XRT, both an absorbed powerlaw plus blackbody model and an absorbed broken powerlaw model are significantly better fits (F-test probability 
$3 \times 10^{-4}$) than a simple absorbed powerlaw. The powerlaw plus blackbody indicates $\NH=1.2 \times 10^{22}$cm$^{-2}$ and a relatively soft photon index of 2.0 with a blackbody temperature of kT$= 0.1$~keV. We note that this value of kT is below the XRT energy band and may therefore indicate a non-physical spectral solution. The absorbed broken powerlaw indicates $\NH=4.3 \times 10^{21}$cm$^{-2}$ and photon indices of $\Gamma_1=1.26$ and $\Gamma_2=2.01$, broken at 3.4~keV. These two models are somewhat degenerate in this dataset, with both models producing a roll over in flux at low (below 0.5~keV) and high (above 50~keV) energies. 

\subsubsection{Segment 8: decay of T$_0$+100~s flare (T$_0$+113 to T$_0$+150~s)}

The decay portion of this flare is well fit by a simple absorbed powerlaw with $\NH=6.0 \times 10^{21}$cm$^{-2}$ and photon index of 2.68. We note, however, that both an absorbed broken powerlaw and absorbed powerlaw plus blackbody are equally good fits to the data.

\subsubsection{Segment 9: rise of T$_0$+160~s flare (T$_0$+159 to T$_0$+171~s)}

The rise of the last flare with overlapping data is well fit by a simple absorbed powerlaw with $\NH=5.4 \times 10^{21}$cm$^{-2}$ and photon index $=2.52$, however the absorbed powerlaw plus blackbody is, strictly, a better fit according to the F-test, though only at about the $80\%$ confidence level (F-test probability = 0.219), with $\NH=5.0 \times 10^{21}$cm$^{-2}$, kT=200~keV and photon index of 2.43. We note that this value of kT is above the XRT-BAT energy band and may therefore indicate a non-physical spectral solution.

\subsubsection{Segment 10: decay of T$_0$+160~s flare (T$_0$+171 to T$_0$+200~s)}

The decay of this flare is well fit by an absorbed powerlaw with $\NH=3.8 \times 10^{21}$cm$^{-2}$ and photon index of 2.55, though as with segment 9, the absorbed powerlaw plus blackbody is also an acceptable fit with $\NH=4.4 \times 10^{21}$cm$^{-2}$, kT=5.2~keV and photon index of 2.83. It should be noted that the BAT flux is very near the noise level in this segment and really provides only an upper limit on the spectral fitting process in the higher energy region.

\subsection{Broadband SED}

We have produced the broadband SED (spectral energy distribution) of the afterglow of GRB~050713A over the time range from T$_0$+20~s to T$_0$+1000~s (Fig 10). This timerange includes detections of the burst afterglow in the optical
from the RAPTOR-S and Liverpool telescopes (corrected for the galactic extinction in this direction of $\AR$=1.04 \citep{Schlegel}) and in the X-ray from {\it Swift} BAT and XRT. It also includes upper limits in the gamma-ray energy range from K-W (whose detectable emission ends
at T$_0$+$\sim$10~s) and in the GeV energy range from MAGIC. A similar SED has been addressed by the MAGIC collaboration in their paper regarding the MAGIC flux upper limit \citep{Albert} in which 
they note that the SED composed of data from {\it Swift} and MAGIC (0.2~keV to 400GeV) is fit by a Band function at low energy and that the MAGIC data are consistent with a single unbroken powerlaw extending from E$_{peak}$ (at $\sim$ 400~keV) to the MAGIC limits up to 500~GeV. We confirm this result, citing a best fit photon index for
a single powerlaw fit from 400~keV to 500~GeV of $\Gamma$ = 2.1 $\pm$ 0.1 and a reduced $\chi^2_r$ = 1.66 for 63 dof. We further note that in performing our fit to the MAGIC data, we have treated the MAGIC upper limits as data points during our fit, thus our photon index or 2.1 is only a lower limit on the true photon index of a powerlaw which would fit the true flux level at GeV energies. Our results here are, therefore, consistent with the analysis of the Albert et. al., in which they show that their data are consistent with a powerlaw photon index of 2.5 from 400~keV to 500~GeV.

We add that a Band function fit is not, however, consistent with the data when we also consider the contemporaneous optical detections. The relative faintness of the optical detections compared
to the X-ray detections combined with the upper limits from K-W and MAGIC requires an absorbed broken powerlaw fit. Figure 10 shows the best fit to the entire SED using 
an absorbed powerlaw (dotted), absorbed broken powerlaw (solid) and absorbed Band function (dashed). The spectral parameters and fit characteristics for each of these fits are shown in Table 7. We have not corrected for the attenuation of flux above 10GeV due to photon-photon interactions with the infrared background \citep{deJager, Kneiske, Primack}, however, our spectral fit results will remain valid independent of this effect due to the constraints placed by the K-W limiting flux measurement from 20~keV to 14~MeV.

\section{Summary}

GRB~050713A is one of the rare bursts observed simultaneously in soft X-rays (XRT), hard X-rays (BAT) and gamma-rays (K-W). 
The broad spectral coverage of these simultaneous measurements has allowed us to fit the early prompt emission, rapid decay, and several flares in the early emission with several different spectral models. In general we find a cutoff powerlaw model to be a good fit to segments with data extending into the MeV range, thus able to constrain the high energy component of the model. For data segments with 0.3--150~keV coverage (BAT and XRT data) we find that a simple absorbed powerlaw is often an adequate fit to the data, though an absorbed powerlaw plus blackbody or absorbed broken powerlaw model seems to sometimes be a marginally better fit during periods of flaring activity. 

The lightcurve structure of GRB~050713A is quite typical of many GRBs that have been observed by \swift. It has an early section showing steep decay slopes of $\alpha > 5$ and 
bright flares extending until T$_0$+$\sim$1~ks, followed by a break to a flatter section with decay slope $\alpha\sim~1.0$ lasting until T$_0$+$\sim$25~ks, followed by a break to a steeper slope of $\alpha=1.45$. 

We have temporally separated the early, flaring portion of the burst into 10 segments and attempted to fit each segment using 4 different spectral models:
1) an absorbed powerlaw 2) an absorbed cutoff or broken powerlaw (cutoff for data extending beyond 150~keV, broken otherwise) 3) an absorbed Band function and 4) an absorbed blackbody plus powerlaw. In all segments where at least two instruments provide significant, simultaneous levels of emission, and hence the spectral data span more than 2 decades in energy,  we find that at
least one of the higher order spectral models is acceptable and, in several cases, is a better fit to the data than a simple absorbed powerlaw. This suggests that the spectral shape of 
GRB flares, while consistent with a simple absorbed powerlaw when viewed through any particular narrow spectral window, is intrinsically fit in the
broadband by a model with attenuated flux above (and possibly below) some threshold energy. 

It has long been known that GRB prompt emission is better
fit by spectral models with a high (and sometimes low) energy cutoff than by a simple absorbed powerlaw \citep{Ryde, Band}, and thus the 
indication that GRB flares are fit by a similar spectral model suggests that similar emission mechanisms may be responsible for the production of flares and for the prompt emission itself, namely internal shocks produced as a result of central engine activity. Since the 
discovery of X-ray flares in GRBs by {\it Swift}, relatively few of the flares have been observed simultaneously across a broad enough energy range to
determine whether such higher order models are necessary to fit their spectra, making the multi-instrument observations of GRB~050713A unique and 
valuable. 

We have also examined the temporal properties of the flares seen in GRB~050713A as early as T$_0$+80~s and as late as T$_0$+10~ks. In all cases we find
the flares to have steep powerlaw rise and decay slopes and 0.1 $<$ $\delta$t/t $<$ 1, which also suggests internal rather than external
shocks as the production mechanism for the flares \citep{Burrows, Ioka}. We have noted that the presence of multiple flares and
the failure of those flares to fit the closure relations associated with the external shock in a wind or constant density ISM further argues in favor of the
internal shock origin for the flares.

We have furthermore discussed the difficulty that {\it Swift} XRT will face in detecting late time flaring 
activity. We have noted that the XRT will have difficulty resolving late, short-timescale flares due to the low XRT count rates 
typically involved. Data from higher throughput instruments such as \xmm {\it EPIC} will be important for constraining flares at these times.
GRB~050713A has simultaneous coverage at moderately late times with \xmm {\it EPIC}, but no conclusive evidence of flaring in the \xmm data has
been found in this case. 

Finally, we have created a broadband SED of the flaring region of GRB~050713A from 0.002~keV to 500GeV at times from T$_0$+20~s to T$_0$+1000~s. We find that the SED is inconsistent with a single absorbed powerlaw or an absorbed Band function and is best fit by an absorbed broken powerlaw. This overall SED again implies that GRB flares are best fit
by a spectral model similar to that of the prompt emission itself and thus suggests a common mechanism for the emission from the prompt phase and from flares.

\acknowledgments
This work is supported at Penn State by NASA contract NAS5-00136.
We gratefully acknowledge the contributions of dozens of members of the {\it Swift} team at PSU, University of Leicester, OAB, GSFC, ASDC, MPE and our subcontractors, who helped make this Observatory possible and to the Flight Operations Team for their support above and beyond the call of duty. This work contains observations obtained with \xmm, an ESA science mission with instruments and contributions directly funded by ESA member states and the USA (NASA). The K-W experiment is supported by Russian Space Agency contract and RFBR grant 06-02-16070.

\clearpage
\begin{deluxetable}{cllcll}
\tablecaption{A Summary of High Energy Observations of GRB
050713A\label{T1}} \tabletypesize{\normalsize} \tablecolumns{5}
\tablewidth{0pt} \tablehead{ \colhead{Observatory} & \colhead{Start
Time } & \colhead{Stop Time } & \colhead{Live-time} &
\colhead{Time Since BAT}\\
\colhead{/Instrument} & \colhead{(UT)} & \colhead{(UT)} &
\colhead{(Seconds)} & \colhead{Trigger (Seconds)} } \startdata
Swift-BAT & 05-07-13-04:29:02.4 & 05-07-13-04:32:00 & 178 & 0\\
Konus-Wind & 05-07-13-04:29:03.1\tablenotemark{*} & 05-07-13-04:37:14.8 & 491.8 & 0.7\\
MAGIC(limit) & 05-07-13-04:29:42 & 05-07-13-05:06:45 & 2223 & 40\\
Swift-XRT & 05-07-13-04:30:14 & 05-08-01-04:37:02 & 167740 & 72\\
XMM-Newton & 05-07-13-10:17:00 & 05-07-13-18:22:00 & 20900 & 21000\\
\enddata
\tablenotetext{*}{The Konus-Wind trigger time corrected for the
propagation time from Wind to Swift}
\end{deluxetable} 
\clearpage

\begin{deluxetable}{cccccc}
\tablecaption{GRB 050713A: X-ray Flares Parameters. \label{T6}}
\tabletypesize{\normalsize} \tablecolumns{6} \tablewidth{0pt}
\tablehead{ \colhead{Start Time} & \colhead{Stop Time} &
\colhead{Duration} &
\colhead{Rise Index $\alpha$\tablenotemark{a}}&
\colhead{Decay Index $\alpha$\tablenotemark{b}} &
\colhead{Peak Flux}\\
\colhead{(s)} & \colhead{ (s)} & \colhead{(s)} &
\colhead{(unitless)} & \colhead{(unitless)} &
\colhead{(ergs~cm$^{-2}$~s$^{-1}$)}} \startdata
79     & 101   & 22 & NA & 5.6 $\pm$ 1.8 & $3 \times 10^{-8}$ (from BAT)\\
101     & 161   & 60 & 23.3 $\pm$ 5 & 8.4 $\pm$ 1.8 & $9 \times 10^{-9}$ \\
161   & 304 & 143 & 8.9 $\pm$ 3 & 6.1 $\pm$ 1.2 & $1.5 \times 10^{-9}$ \\
9751 & 11840  & 2089 & 5.76 $\pm$ 1.8 & 11.0 $\pm$ 2.4 & $1 \times 10^{-11}$ \\
\enddata
\tablenotetext{a}{\begin{footnotesize}Index $\alpha$ of a powerlaw fit to the rise of the flare with T$_0$=BAT trigger time; $\Gamma_{\nu} \propto ($t-T$_0)^{\alpha}$\end{footnotesize}}
\tablenotetext{a}{\begin{footnotesize}Index $\alpha$ of a powerlaw fit to the decay of the flare with T$_0$=BAT trigger time; $\Gamma_{\nu} \propto ($t-T$_0)^{-\alpha}$\end{footnotesize}}
\end{deluxetable}
\clearpage

\begin{deluxetable}{ccccc}
\tablecaption{MAGIC upper limit (95\% CL) on GRB~050713A
between T$_0+40$~s and T$_0+130$~s.
Limits include a systematic uncertainty of 30\%.
1 C.U. (\emph{Crab Unit}) =
$1.5 \times 10^{-6} \times (\mathrm{E/GeV})^{-2.58}
\:\mathrm{ph}$~cm$^{-2}$~s$^{-1}$~GeV$^{-1}$. \label{T7}}
\tabletypesize{\normalsize} \tablecolumns{5} \tablewidth{0pt}
\tablehead{ \colhead{Energy} & \colhead{Excess evts.} &
\colhead{Eff. Area} & \colhead{Flux lim} & \colhead{Flux lim}\\
\colhead{(GeV)} & \colhead{(uplim)} & \colhead{($\times 10^8$cm$^2$)} &
\colhead{(cm$^{-2}$~keV$^{-1}$s$^{-1}$)} & \colhead{(C.U.)} }
\startdata
$175-225$  & $\phantom08.5$ & 1.7 & $1.3 \times 10^{-17}$ & 7.6\\
$225-300$  &          10.4  & 3.4 & $3.9 \times 10^{-18}$ & 4.8\\
$300-400$  & $\phantom06.0$ & 5.3 & $1.6 \times 10^{-18}$ & 3.8\\
$400-1000$ & $\phantom04.3$ & 6.5 & $2.3 \times 10^{-19}$ & 3.3\\
\enddata
\end{deluxetable}
\clearpage

\begin{deluxetable}{cccc}
\tablecaption{GRB 050713A: Ground Based Optical and Radio Followup. \label{T8}}
\tabletypesize{\normalsize} \tablecolumns{6} \tablewidth{0pt}
\tablehead{\colhead{Observatory} &  \colhead{Time} & \colhead{Band} &
\colhead{Magnitude/Limit}} \startdata
McDonald Obs, Tex & T$_0$+22.4~s     & unfilt   & 17.7 (lim) \\
RAPTOR-S, LANL & T$_0$+99.3~s     & R    & 18.4 $\pm$ 0.18 \\
Liverpool Robotic Telescope, Canary Islands & T$_0$+180~s   & r' & 19.2 \\
\swift & T$_0$+252~s  & V & 17.98 \\
\swift & T$_0$+309~s  & U & 17.81 \\
\swift & T$_0$+311~s  & UVM2 & 17.13 \\
\swift & T$_0$+325~s  & UVW1 & 16.85 \\
\swift & T$_0$+326~s  & UVW2 & 17.08 \\
\swift & T$_0$+351~s  & B & 18.08 \\
Red Buttes Obs, Wy & T$_0$+27m & R  & 19.4 (lim) \\
Red Buttes Obs, Wy & T$_0$+31m     & I  & 18.2 (lim) \\
Nordic Optical Tel & T$_0$+47m     & R  & $<$ DSS limit \\
Galileo National Telescope, Canary Islands &  T$_0$+48m   & I    & $<$ DSS limit \\
ARC Telescope, Apache Point Obs & T$_0$+53m & J,H,K  & detected \\
Red Buttes Obs, Wy & T$_0$+93m     & R   & 19.4 (lim) \\
Red Buttes Obs, Wy & T$_0$+98m     & I   & 18.7 (lim) \\
Lulin Telescope, Taiwan & T$_0$+10.3h   & R    & 22.4 (lim) \\
VLA, NRAO & T$_0$+4.3d &8.5 GHz    & 96 microJan \\
\enddata
\end{deluxetable}
\clearpage

\begin{deluxetable}{cccc}
\tablecaption{Swift and \xmm spectral fits pre-break and post-break. \label{T7}}
\tabletypesize{\normalsize} \tablecolumns{6} \tablewidth{0pt}
\tablehead{\colhead{Observatory} & \colhead{photon index} & \colhead{\NH~(cm$^{-2}$)} & \colhead{comment}} \startdata
\swift orbit 1 & 2.28 $\pm$ 0.04  & $4.8 \pm 0.2 \times 10^{21}$ & pre- energy injection phase\\
\swift orbit 2  & 1.90 $\pm$ 0.13 & $3.1 \pm 0.4 \times 10^{21}$ & energy injection phase\\
\swift orbit 3  & 2.25 $\pm$ 0.23 & $4.1 \pm 0.7 \times 10^{21}$ & flare during energy injection\\
\swift after orbit 3  & 2.8 $\pm$ 0.6 & $5.6 \pm 0.2 \times 10^{21}$ & post-break\\
\xmm & 2.1 $\pm$ 0.05 & $3.0 \pm 0.1 \times 10^{21}$ & post-break\\
\enddata
\end{deluxetable}
\clearpage

\begin{deluxetable}{ccccc|c|ccccc}
\tablecaption{GRB 050713A: Joint Spectral Fits - We group the data into segments to separate times which may show different spectra. Segments 1-4 contain BAT and K-W data and are segmented to separate prompt emission from the rapid decay phase. Segment 5 contains BAT data only and segments 6-10 contain XRT and BAT data. These are segmented to separate the rise and decay of each flare. We attempt fits to each segment using 4 spectral models: 1) an absorbed powerlaw 2) an absorbed cutoff or broken powerlaw 3) an absorbed Band function and 4) an absorbed blackbody plus powerlaw. In segments where a particular model was inapplicable or the fit did not converge, NA is entered in the table. \label{T6}}
\tabletypesize{\normalsize} \tablecolumns{11} \tablewidth{0pt}
\tablehead{\colhead{segment} & \colhead{\textbf{1}} & \colhead{\textbf{2}} & \colhead{\textbf{3}} & \colhead{\textbf{4}} & \colhead{\textbf{5}} & \colhead{\textbf{6}} & \colhead{\textbf{7}} & \colhead{\textbf{8}} & \colhead{\textbf{9}} & \colhead{\textbf{10}}} \startdata
$\delta$t~(s)& \begin{footnotesize}--70~-~--49.5\end{footnotesize} & \begin{footnotesize}0-8.5\end{footnotesize} & \begin{footnotesize}8.5-25\end{footnotesize} & \begin{footnotesize}0-16.5\end{footnotesize} & \begin{footnotesize}59-68\end{footnotesize} & \begin{footnotesize}68-95\end{footnotesize} & \begin{footnotesize}100-113\end{footnotesize} & \begin{footnotesize}113-150\end{footnotesize} & \begin{footnotesize}159-171\end{footnotesize} & \begin{footnotesize}171-200\end{footnotesize}\\
\hline
Instr & BAT & and & K-W & &BAT& & XRT & and & BAT & \\
\hline
pl:\NH & NA & NA & NA & NA & NA & 0.52 & 0.59 & 0.60 & 0.54 & 0.38\\
pl:PhInd & 1.26 & 1.44 & 1.61 & 1.47 & 1.83 & 2.47 & 1.72 & 2.68 & 2.52 & 2.55\\
pl:$\chi_{\nu}^2$ & 1.54 & 1.96 & 1.39 & 1.95 & 0.98 & 1.20 & 1.21 & 1.07 & 1.04 & 1.00\\
pl:dof & 12 & 101 & 91 & 98 & 26 & 74 & 98 & 125 & 22 & 37\\
\hline
cutoffpl:PhInd  & 0.913 & 1.26 & 1.31 & 1.25 & NA & NA & NA & NA & NA & NA\\
cutoffpl:E$_{peak}$  & 270 & 421  & 312 & 400 & NA & NA & NA & NA & NA & NA\\
cutoffpl:$\chi^2$  & 1.20 & 1.60 & 1.02 & 1.29 & NA & NA & NA & NA & NA & NA\\
cutoffpl:dof & 11 & 110 & 90 & 100 & NA & NA & NA & NA & NA & NA\\
\hline
bknpl:\NH & NA & NA & NA & NA & NA & 0.58 & 0.43 & 0.45 & NA & NA\\
bknpl:PhInd1 &  NA & NA & NA & NA & 0.95 & 2.69 & 1.26 & 1.93 & NA & NA\\
bknpl:E$_{break}$ &  NA & NA & NA & NA & 5.8 & 4.1 & 3.4 & 1.8 & NA & NA\\
bknpl:PhInd2 &  NA & NA & NA & NA & 2.1 & 1.5 & 2.0 & 2.8 & NA & NA\\
bknpl:$\chi_{\nu}^2$ &  NA & NA & NA & NA & 1.06 & 1.31 & 1.03 & 1.04 & NA & NA\\
bknpl:dof &  NA & NA & NA & NA & 23 & 73 & 97 & 124 & 21 & 36\\
\hline
Band:\NH & NA & NA & NA & NA & NA & 0.33 & 0.49 & 0.54 & 0.41 & 0.33\\
Band:$\alpha$ & -1.13 & -1.27 & -1.27 & -1.29 & -2.01 & -1.69 & -1.46 & -2.32 & -1.98 & -2.17\\
Band:$\beta$ & -1.26 & -9.36 & -2.39 & -9.29 & -9.07 & -1.88 & -9.36 & -9.32 & -9.04 & -8.78\\
Band:E$_{peak}$ & 101 & 761 & 244 & 636 & 994 & 10.6 & 30.7 & 10.6 & 10.6 & 10.6\\
Band:$\chi_{\nu}^2$ & 1.85 & 1.61 & 1.03 & 1.28 & 1.06 & 2.11 & 1.11 & 1.13 & 1.26 & 1.12\\
Band:dof & 10 & 109 & 89 & 101 & 23 & 73 & 97 & 124 & 21 & 36\\
\hline
pl+bb:\NH & NA & NA & NA & NA & NA & 0.59 & 1.17 & 0.63 & 0.50 & 0.44\\
pl+bb:kT & 3.59 & 59.4 & 26.4 & 44.97 & NA & 7.1 & 0.1 & 4.4 & 200 & 5.2\\
pl+bb:bb$_{norm}$ & 0.068 & 2.74 & 0.46 & 1.47 & NA & 0.07 & 2.48 & 0.01 & 2.90 & 0.01\\
pl+bb:PhInd & 1.00 & 1.64 & 1.76 & 1.63 & NA & 2.75 & 2.00 & 2.77 & 2.43 & 2.83\\
pl+bb:pl$_{norm}$ & 0.053 & 13.97 & 7.69 & 11.65 & NA & 1.53 & 2.32 & 2.03 & 0.51 & 0.44\\
pl+bb:$\chi_{\nu}^2$ & 1.15 & 1.79 & 1.03 & 1.60 & NA & 1.33 & 1.04 & 1.07 & 0.90 & 0.98\\
pl+bb:dof & 10 & 109 & 89 & 101 & NA & 73 & 97 & 124 & 21 & 36\\
\enddata
\end{deluxetable}

\begin{deluxetable}{lc}
\tablecaption{GRB 050713A: SED Fit Data - A broadband SED (R-band optical data points
to 500GeV upper limits) has been created and we show the result of fits of
3 spectral models: 1) an absorbed powerlaw 2) an absorbed broken powerlaw and 3) an absorbed
Band function. We do not report errors on the parameters in the powerlaw fit or Band function because these models are
clearly unacceptable as shown by the large values of $\chi_{\nu}^2$. Only the broken powerlaw is an
acceptable fit to the entire SED.  \label{T8}}
\tabletypesize{\normalsize} \tablecolumns{6} \tablewidth{0pt}
\tablehead{\colhead{Model:Param} & \colhead{Value}}
 \startdata
pl:\NH & $4.3 \times 10^{21}$ cm$^{-2}$\\
pl:PhInd & 2.14\\
pl:$\chi_{\nu}^2$ & 10 (65 dof)\\
\hline
bknpl:\NH & $2.9 \pm 0.3 \times 10^{21}$ cm$^{-2}$\\
bknpl:PhInd1 & 1.1 $\pm$ 0.1 \\
bknpl:E$_{break}$ & 1.3 $\pm$ 0.2 keV\\
bknpl:PhInd2 & 2.2 $\pm$ 0.1 \\
bknpl:$\chi_{\nu}^2$ & 1.20 (63 dof) \\
\hline
Band:\NH & $2.1 \times 10^{21}$ cm$^{-2}$\\
Band:$\alpha$ & --1.3 \\
Band:$\beta$ & --2.2 \\
Band:E$_{peak}$ & 10.6 keV\\
Band:$\chi_{\nu}^2$ & 2.97 (63 dof) \\
\enddata
\end{deluxetable}

\clearpage

\onecolumn
\begin{figure}[ht]
\figurenum{1} \epsscale{1.0} \plotone{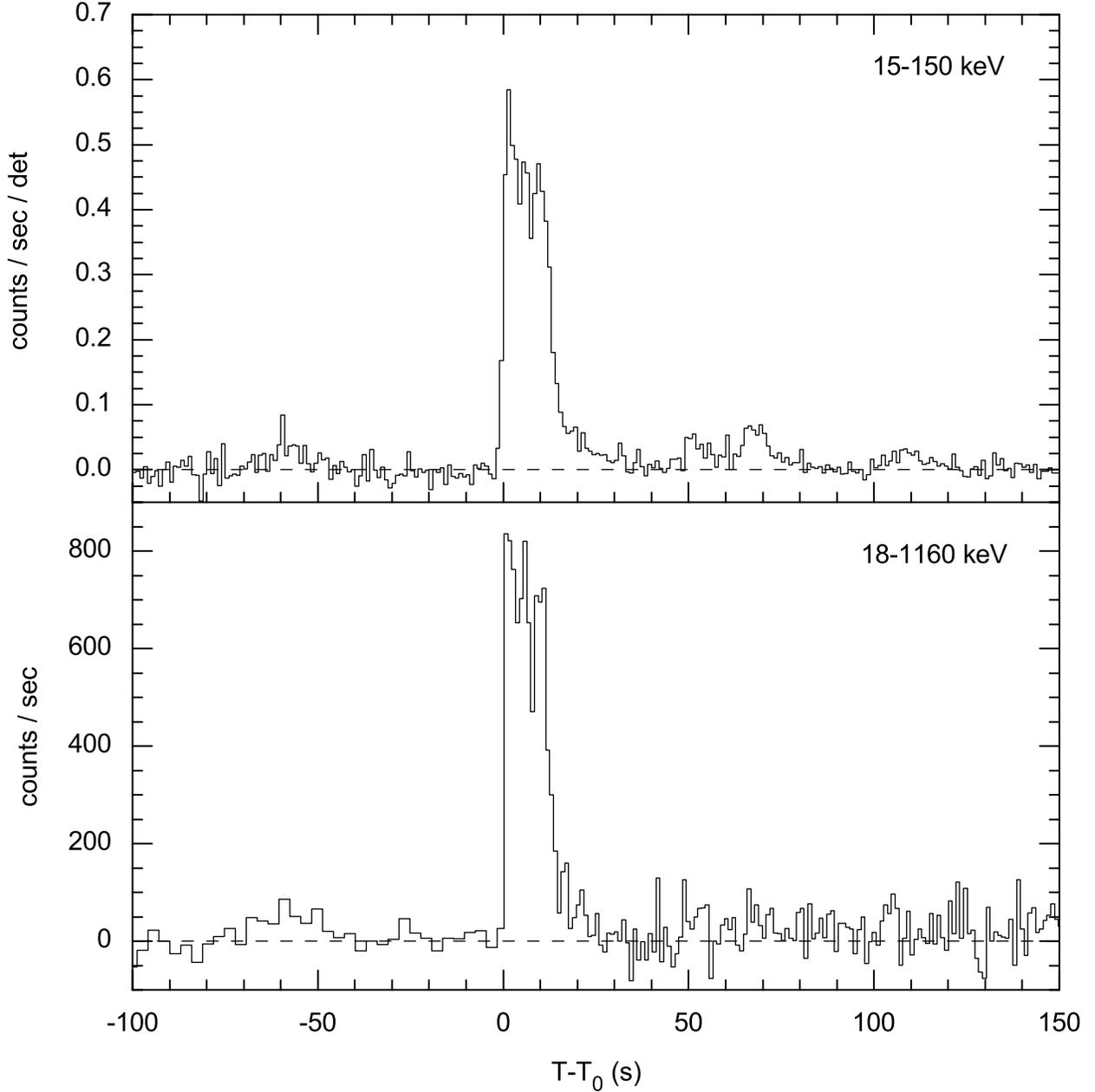} \caption{Background 
subtracted BAT (top panel) and Konus-WIND (bottom) light
curves on the same time scale. The plots have been adjusted 
so that the trigger time for both plots is the same relative to the burst. 
T$_0$ in the lower plot is T$_0$(BAT) 
plus the propagation time between the spacecrafts (0.742~s). BAT data are binned to 1~s resolution
throughout. K-W data are binned to 2.94~s resolution in survey mode prior to the burst trigger and 
are binned to 1~s resolution in GRB follow-up mode after the trigger. Note that the precursor at T$_0$--65~s
is detected in both BAT and K-W while post-trigger flares seen in the BAT at T$_0$+50~s, T$_0$+65~s and T$_0$+105~s 
are not clearly detected by K-W. This suggests a harder spectrum for the precursor than the post-trigger flares,
which is confirmed by joint BAT/K-W spectral fits. The main burst consists of 3 closely spaced, overlapping pulses
in both the BAT and K-W energy ranges. The K-W lightcurve decays rapidly to background level by T$_0$+15~s while
the BAT lightcurve continues to show low level emission out to T$_0+\sim$200~s.}.
\label{coolplot}
\end{figure}
\clearpage

\onecolumn
\begin{figure}[ht]
\figurenum{2} \epsscale{1.0} \plotone{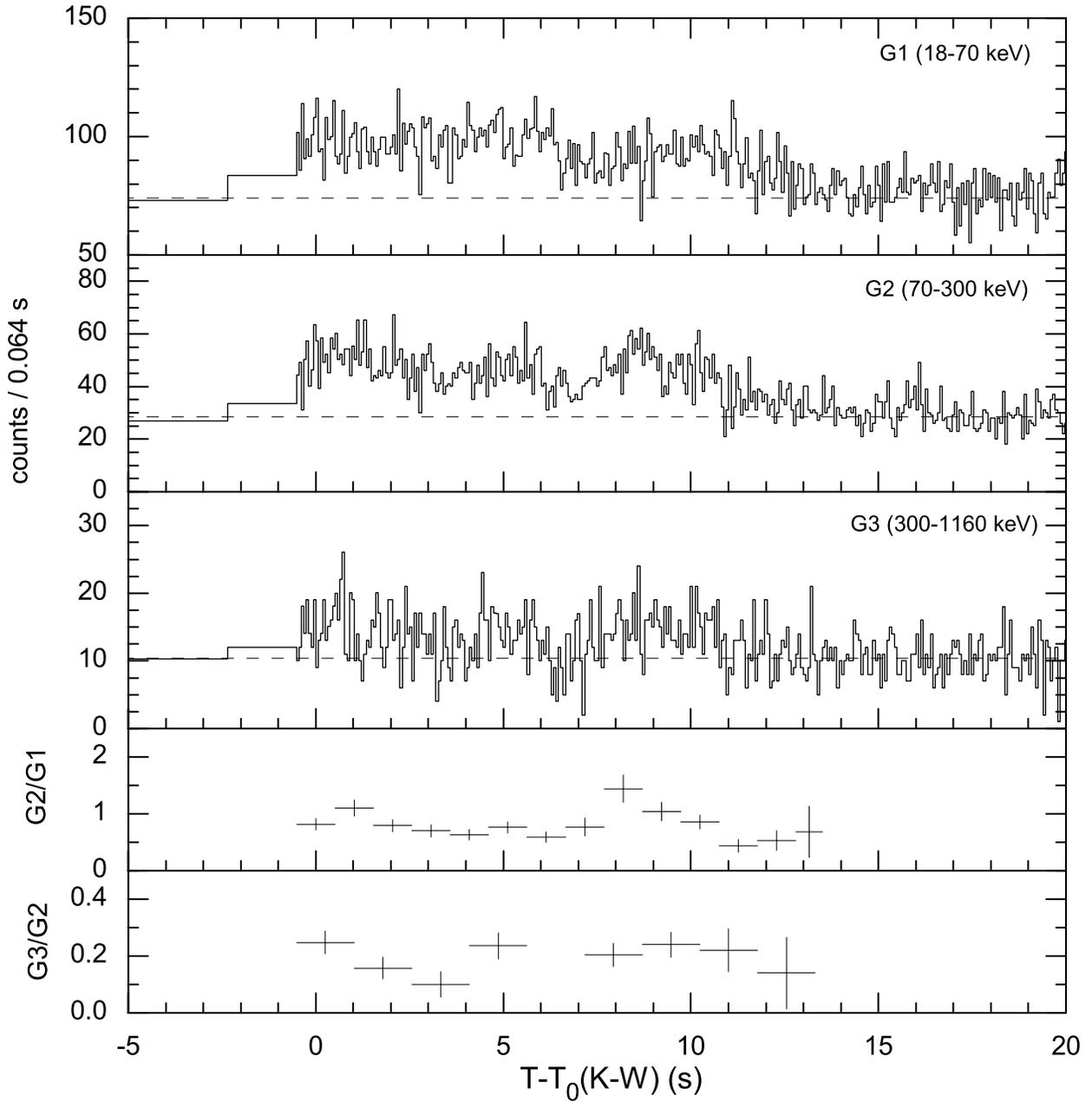} \caption{Plot of
Konus-Wind data in 3 bands and associated band ratios during burst prompt emission. Data binning is 64ms.}.
\label{coolplot}
\end{figure}
\clearpage

\onecolumn
\begin{figure}[ht]
\figurenum{3} \epsscale{1.0} \plotone{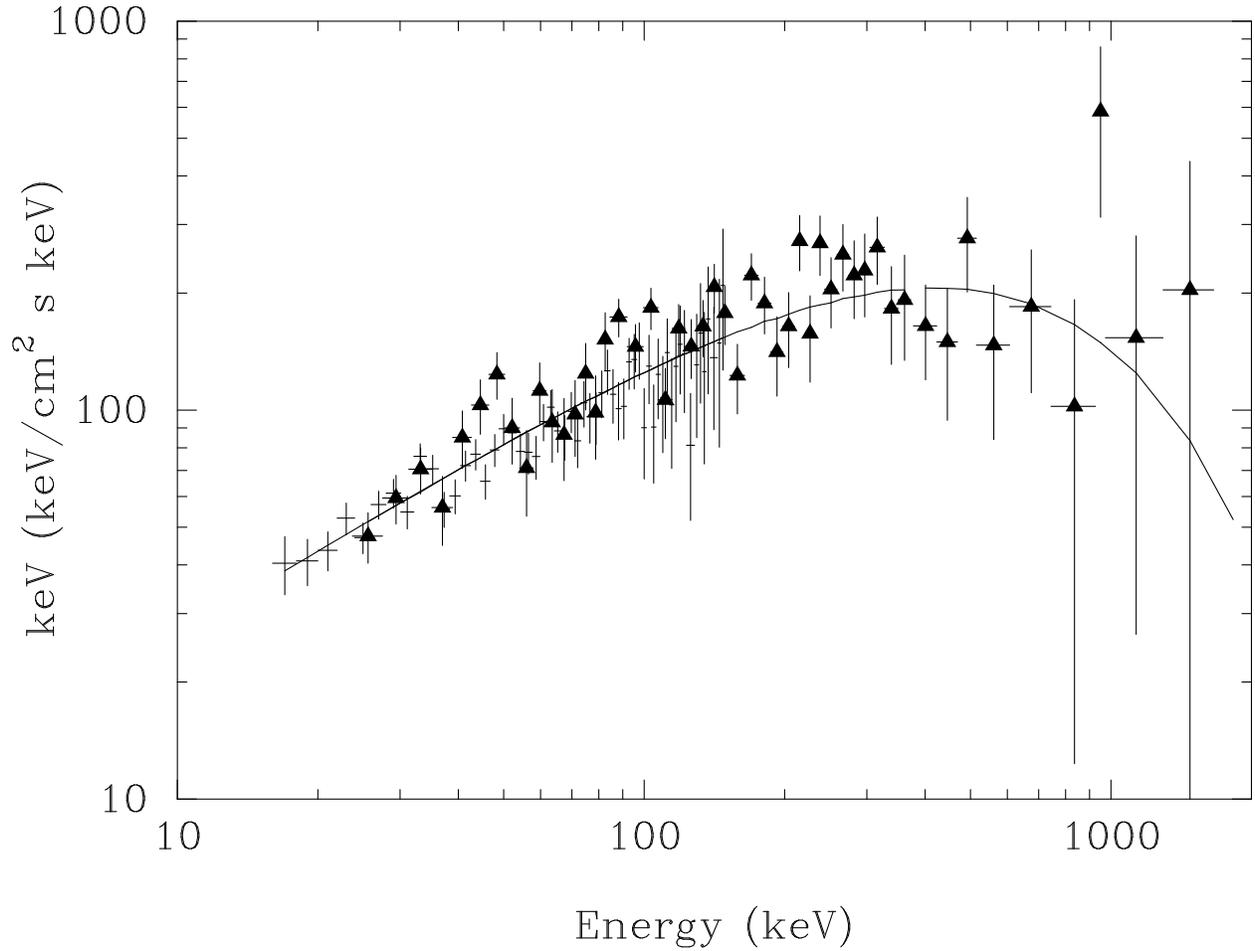} \caption{Plot of
joint spectral energy distribution of Konus-Wind and BAT data during burst prompt emission, showing E$_{peak}$ = 421~keV. 
K-W data are filled triangles, BAT data are crosses. Data channels have been grouped where appropriate to produce significant data points.}.
\label{coolplot}
\end{figure}
\clearpage

\begin{figure}[ht]
\figurenum{4} \epsscale{1.0} \plotone{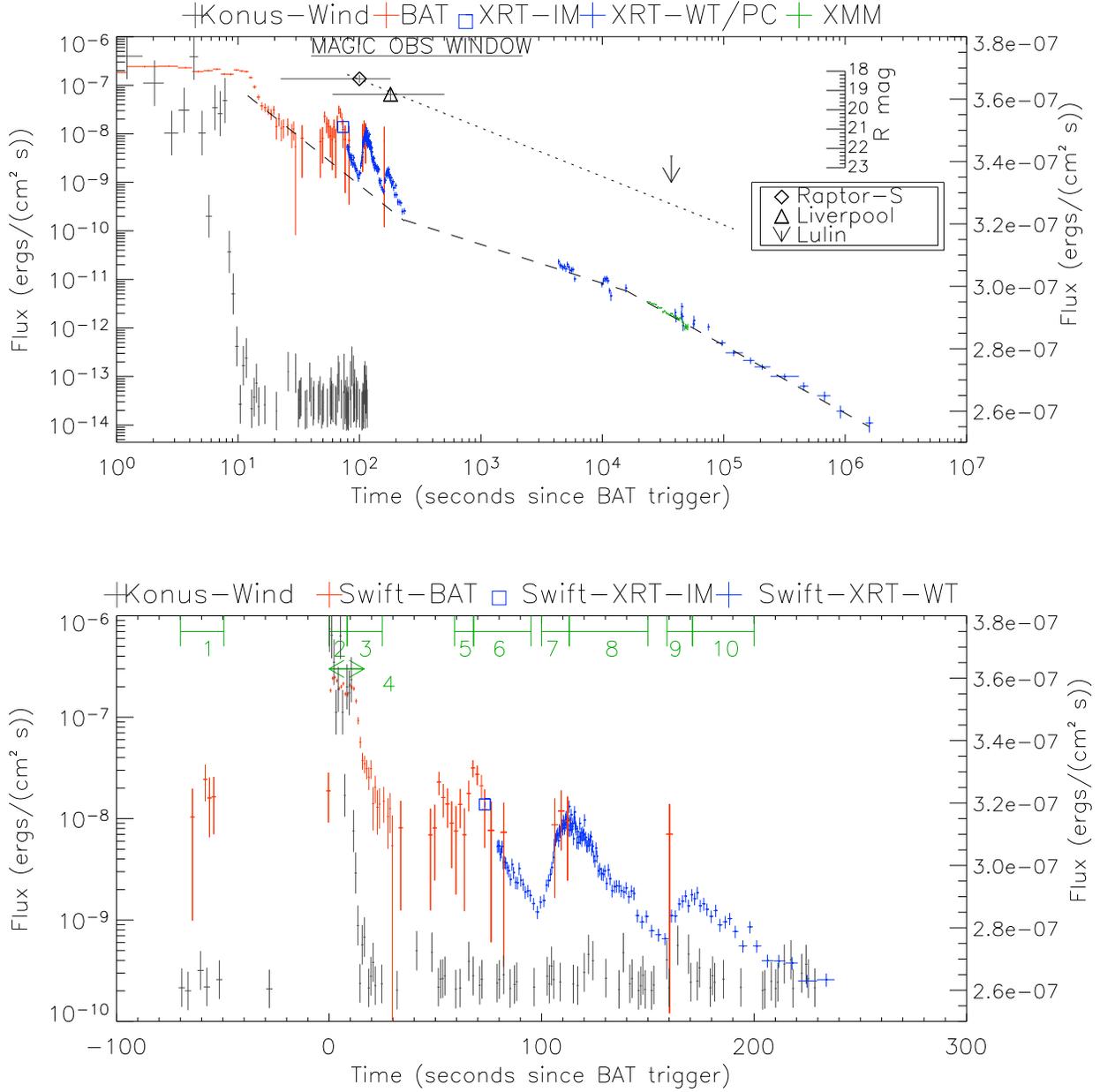}
\caption{X-ray/gamma-ray/optical lightcurve of GRB~050713A. Top: multicolored points are {\it Swift} 
and XMM data scaled to the left Y-axis. Black crosses are K-W data scaled 
to the right Y-axis. Fluxes are extrapolated into the 0.2--10~keV energy range. 
The diamond, cross and arrow are optical observations and scaled 
to the inset Y-axis. The scaling of
the inset Y-axis is consistent with the outer, left Y-axis such that 1 magnitude is equal to a factor of 2.5 in flux. 
The window of MAGIC observations is shown by the horizontal bar.
The dashed line is the supposed underlying powerlaw decay.
Data from T$_0$+4~ks to T$_0$+16~ks
are well fit by a flatter powerlaw of slope t$^{-0.8}$, implying an energy injection phase. 
A break to a steeper decay of t$^{-1.45}$ occurs at T$_0$+$\sim$25~ks. 
We note the similar decay 
slopes in each of the three flares seen by XRT. Optical data are plotted with a
fitted powerlaw decay of t$^{-1.0}$. Bottom: a close-up of the flares. 
Green bars indicate the segments of joint spectral fits.} \label{xrt_lightcurve}
\end{figure}
\clearpage

\begin{figure}[ht]
\figurenum{5} \epsscale{1.0} \plotone{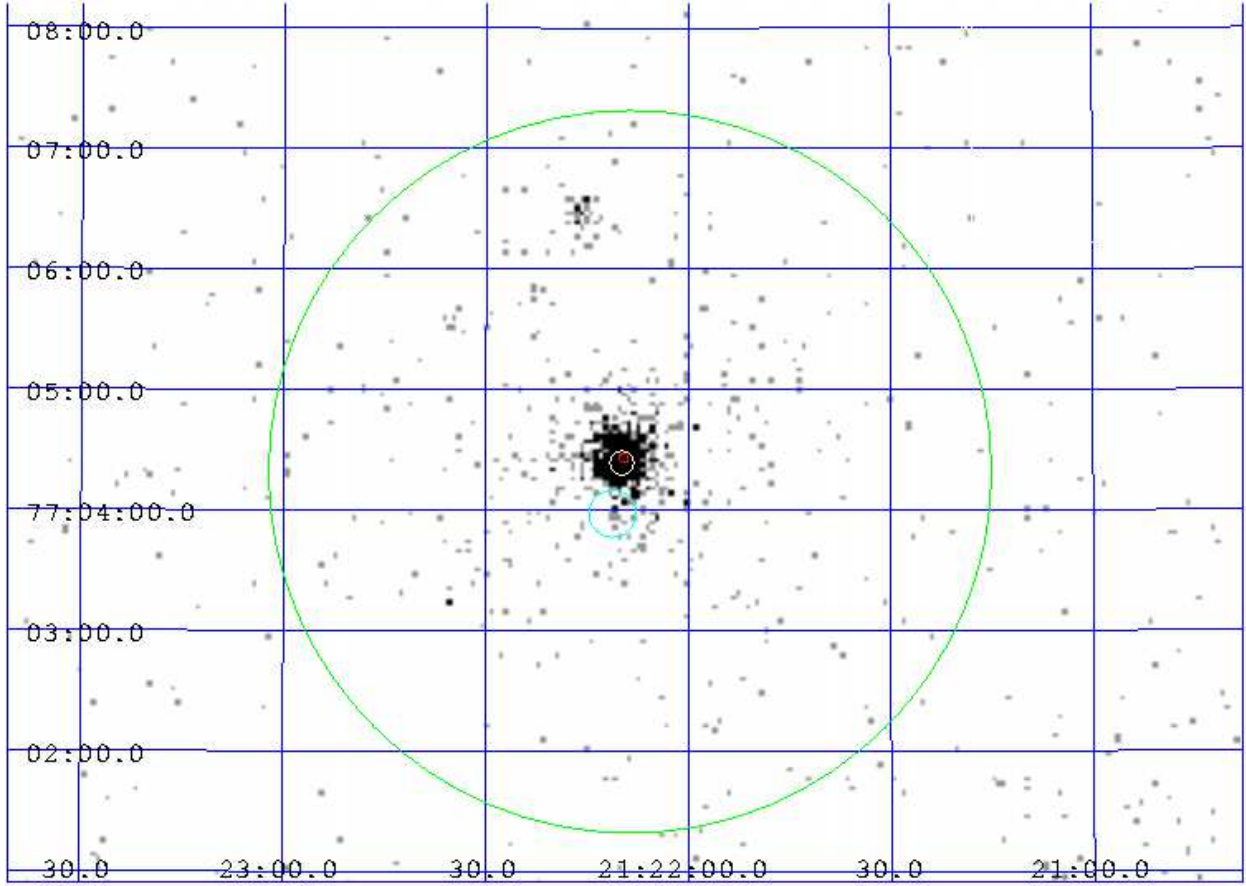}
\caption{XRT image with BAT and XRT optical error circles plotted. Green~=~BAT; White=XRT; Red=optical.
The light blue circle indicates the location of the serendipitous source located 30 arcseconds south of
the GRB which has been subtracted from the data.}
\label{xrt image}
\end{figure}
\clearpage

\begin{figure}[ht]
\figurenum{6} \epsscale{1.0} \plotone{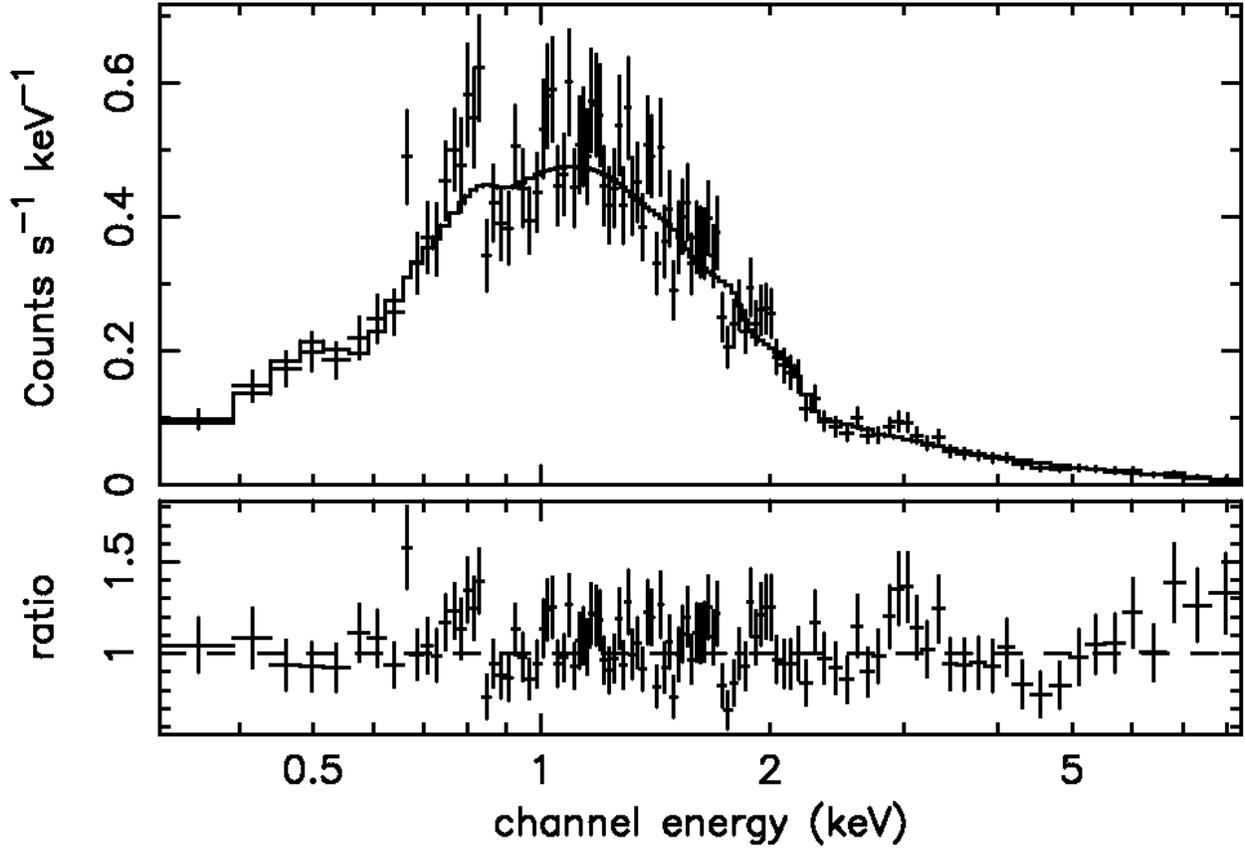}
\caption{PN spectrum from the first 8~ks of the \xmm\ observation.
The top panel shows the PN data (crosses) with best fit model (solid line)
overlaid, which consists of an absorbed power-law with photon index $=2.07$
and $\NH=3.2 \times 10^{21}$\,cm$^{-2}$. The bottom panel shows the
data/model ratio residuals to this continuum model. A weak excess of counts
is seen near 0.8~keV and 3~keV, although if interpreted as emission lines,
the detection is not significant.}
\label{xrt image}
\end{figure}
\clearpage

\onecolumn
\begin{figure}[ht]
\figurenum{7} \epsscale{1.0} \plotone{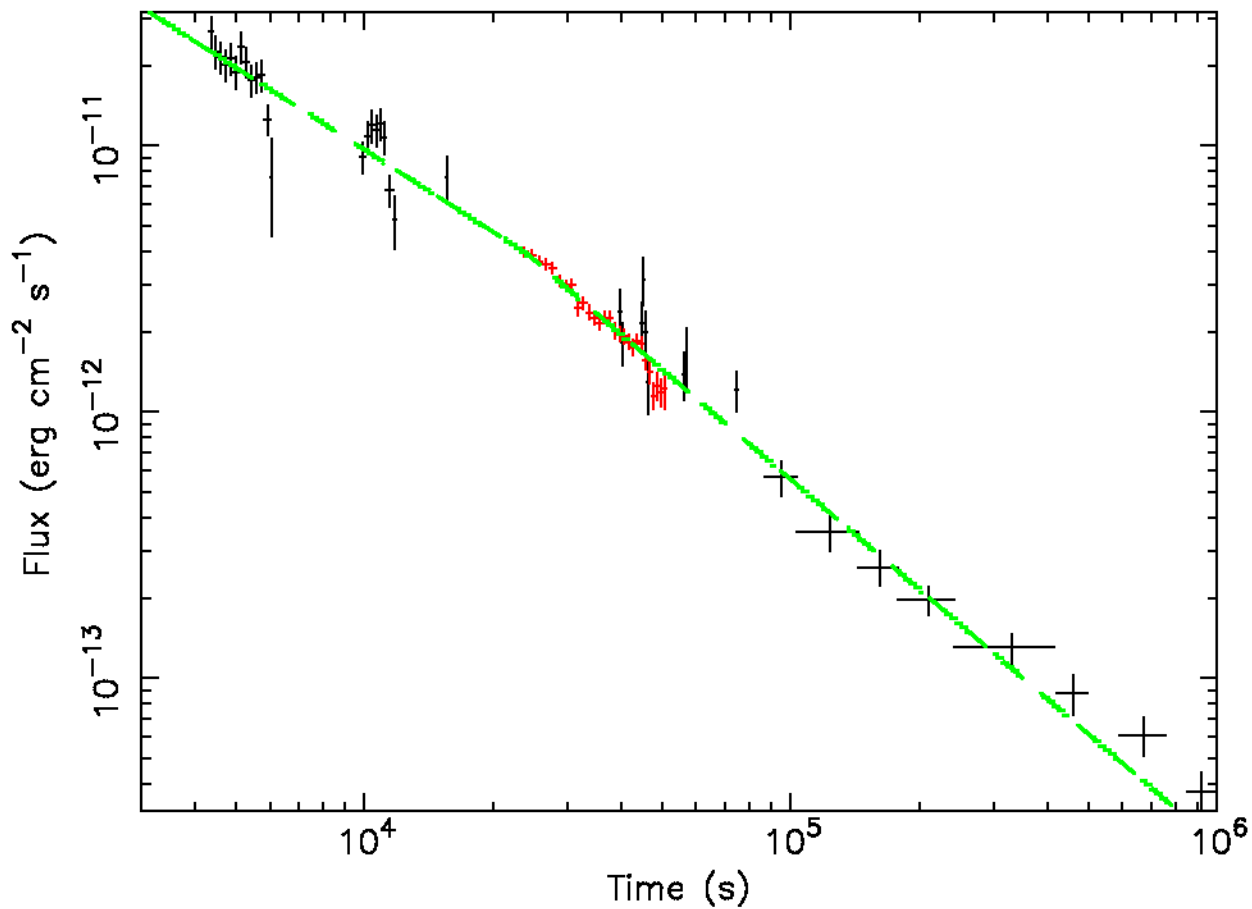}
\caption{Joint \swift\ XRT and \xmm\ PN lightcurve. \swift data are from T$_0$+4~ks to T$_0$+1000~ks.
\swift\ XRT points are shown in black and \xmm\ as red. The
afterglow flux is measured in the 0.5--10~keV band, not correcting for absorption.
The solid line plotted to the different segments of data is a
broken power-law decay model, outlined in the text. The \xmm\ decay
index ($\alpha=1.45$) is considerably steeper than in the XRT at
earlier times ($\alpha=1.0$), suggesting
that a break occurs in the lightcurve decay at around T$_0$+25~ks.}
\label{f7}
\end{figure}
\clearpage

\begin{figure}[htp]
\figurenum{8}
\centering
\includegraphics[width=.75\linewidth]{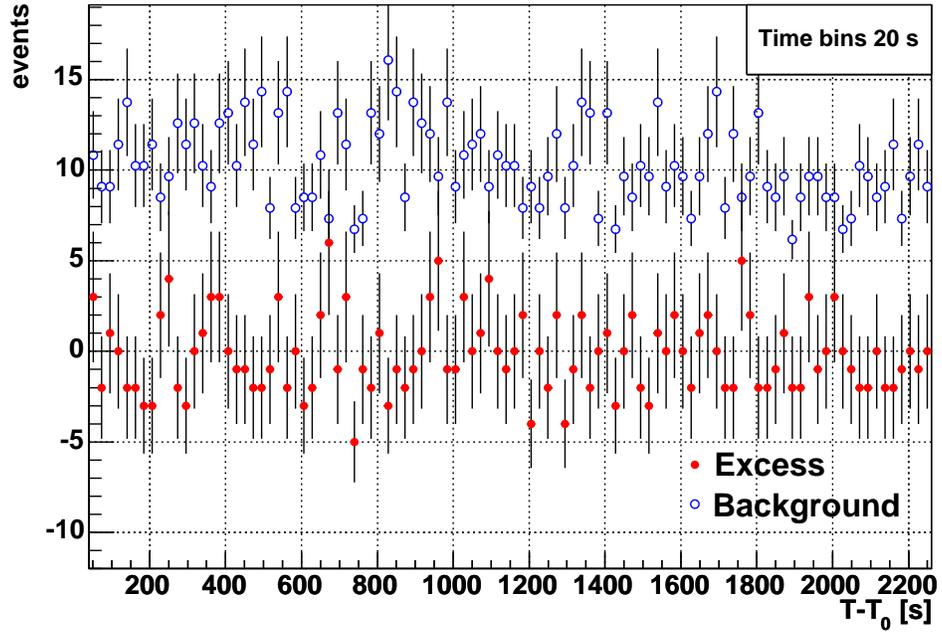}
\caption[Excess Events of MAGIC]{%
\label{fig:excessevents}
MAGIC Observations. Filled circles: number of excess events for 20~s intervals, in the 37 min window after the burst onset.
Open circles: number of background events in the signal region. No significant source signal is detected above the background.}
\end{figure}
\clearpage

\begin{figure}[ht]
\figurenum{9} \epsscale{1.0} \plotone{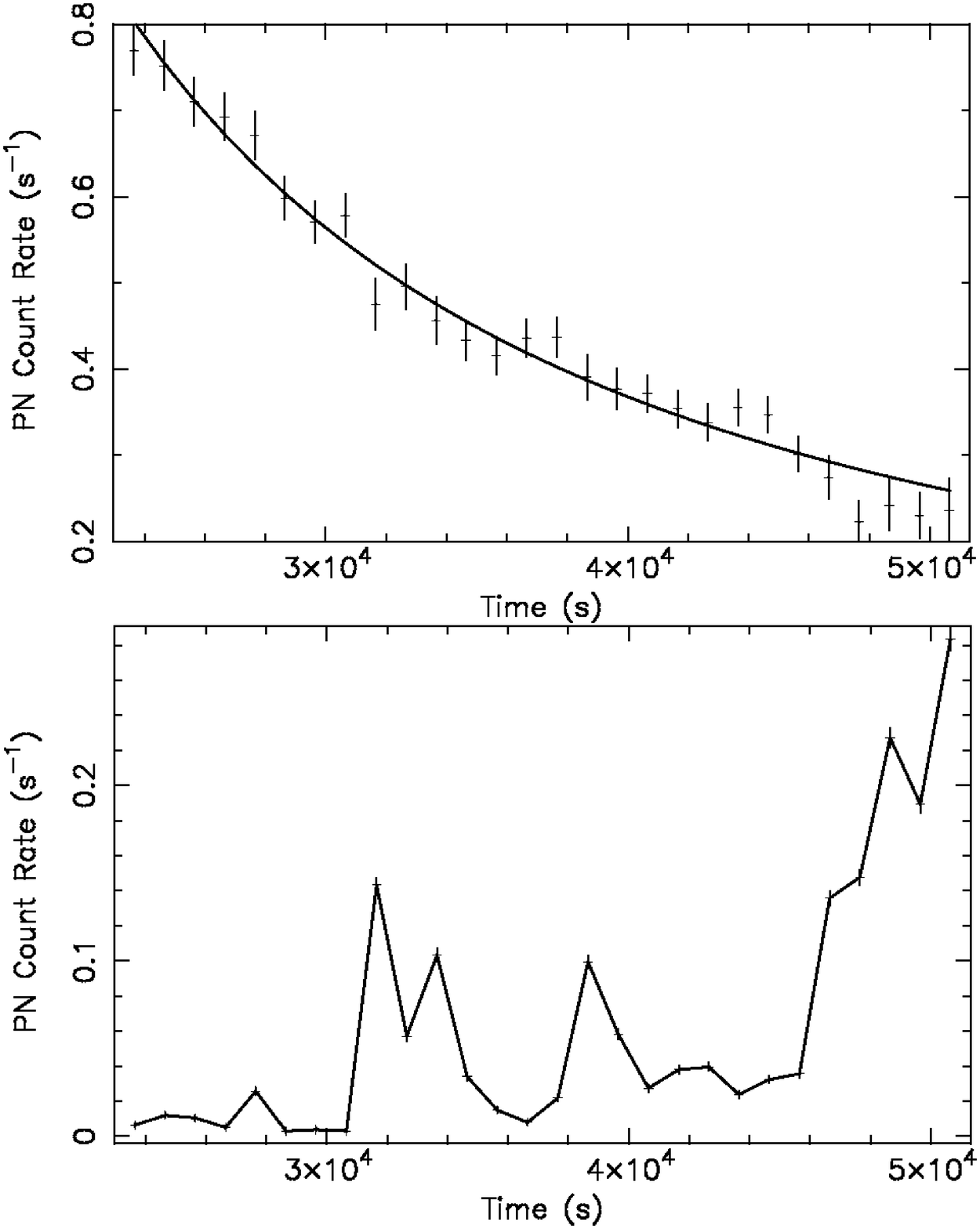}
\caption{\xmm\ lightcurves for the afterglow of GRB 050713A. The top
panel shows the background subtracted afterglow lightcurve for the PN
detector. Crosses show the GRB source counts ($1\sigma$ errors), the solid
line shows the best fit decay rate of t$^{-1.45}$. Time is plotted compared
to the initial BAT trigger. The bottom panel shows
the background lightcurve for the PN, normalized to the size of the source
extraction region for comparison.}
\label{f9}
\end{figure}
\clearpage

\begin{figure}[ht]
\figurenum{10} \epsscale{1.0} \plotone{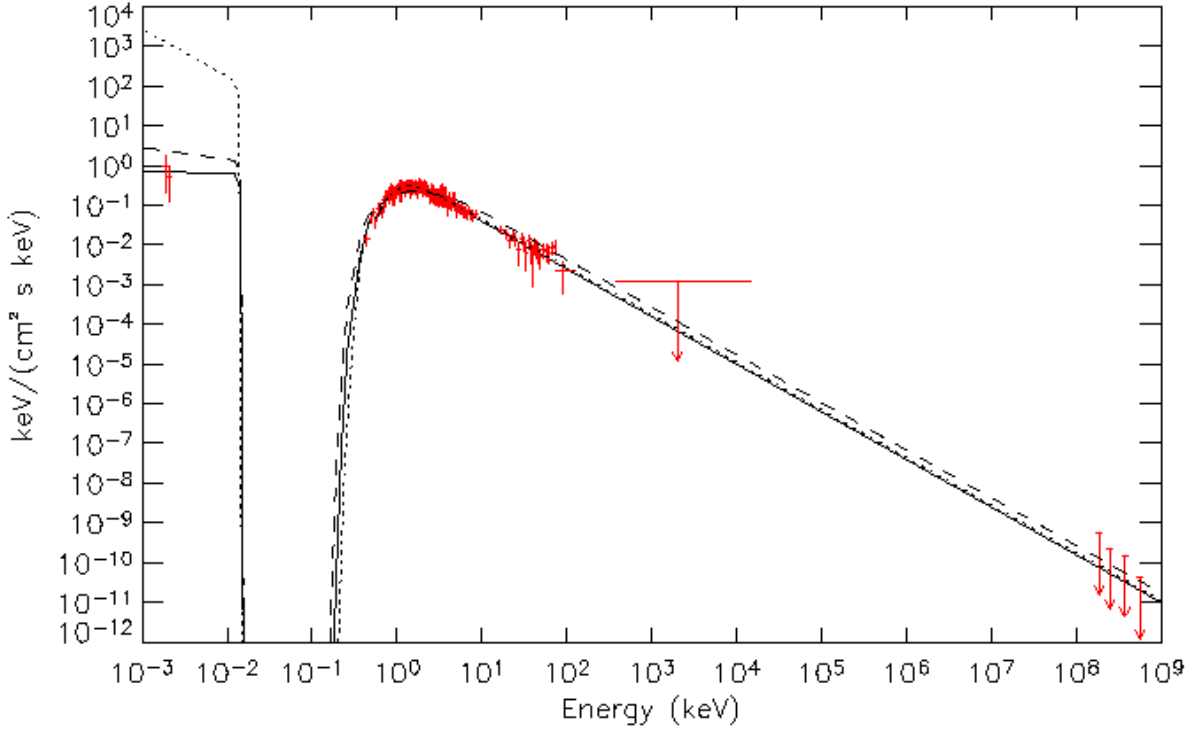}
\caption{Combined multi-platform SED of the early afterglow of GRB~050713A from T$_0$+20~s to T$_0$+1000~s.
Optical data are from RAPTOR-S at LANL and the Liverpool robotic telescope, soft X-ray (0.2--10~keV)
data are from {\it Swift} XRT, hard X-ray (15--150~keV) data are from {\it Swift} BAT and gamma-ray
upper limits are from Konus-Wind (0.5--14~MeV) and MAGIC (175--500GeV). The three lines plotted over the
data represent the 3 models discussed as proposed fits to the SED in the text. The absorbed broken
power law (solid) is the only acceptable fit. The absorbed Band function (dashed) and simple absorbed
powerlaw (dotted) do not appear reconcilable with the data. The results suggest that the GRB flare emission
is characterized by a single mechanism well represented by a broken powerlaw, or that a more
complex, possibly multi-component emission mechanism is required to explain the complete SED.}
\label{f10}
\end{figure}
\clearpage

\end{document}